\documentclass{ejm}%
\usepackage{amsmath}
\usepackage{amsfonts}
\usepackage{amssymb}
\usepackage{graphicx}%
\usepackage[normal]{subfigure}
\newtheorem{theorem}{Theorem}

\newtheorem{lemma}[theorem]{Lemma}

\newtheorem{proposition}[theorem]{Proposition}

\newcommand{\eps}{{\displaystyle \varepsilon}}
\def\goesto{\rightarrow}

\newcommand{\bsub}{\begin{subequations}}
\newcommand{\alp}{\alpha}
\newcommand{\esub}{\end{subequations}$\!$}
\renewcommand{\theequation}{\arabic{section}.\arabic{equation}}
\renewcommand{\thesection}{\arabic{section}}
\pubyear{2005}
\volume{000}
\pagerange{\pageref{firstpage}--\pageref{lastpage}}
\affiliation{
$~^\star$
Department of Mathematics, Dalhousie University, Halifax, N.S., Canada\\
tkolokol\symbol{64}gmail.com\\
$~^\dag$
Department of Mathematics, University of British Columbia, Vancouver, B.C., Canada\\
ward\symbol{64}math.ubc.ca\\
$~^\ddag$\ Department of Mathematics,
The Chinese University of Hong Kong,
Shatin, Hong Kong \\
wei\symbol{64}math.cuhk.edu.hk
}

\begin{document}

\title{Self-Replication of Mesa Patterns in Reaction-Diffusion
Systems} \author{T.\ns K\ls O\ls L\ls O\ls K\ls O\ls L\ls N\ls I\ls
K\ls O\ls V$~^{\star}$, \ns M.\ls J.\ns W\ls A\ls R\ls D$~^{\dag}$ \ns
J.\ns W\ls E\ls I$~^{\ddag}$ } \maketitle

\begin{abstract}
Certain two-component reaction-diffusion systems on a finite interval
are known to possess mesa (box-like) steady-state patterns in the
singularly perturbed limit of small diffusivity for one of the two
solution components. As the diffusivity $D$ of the second component is
decreased below some critical value $D_c$, with $D_c=O(1)$, the
existence of a steady-state mesa pattern is lost, triggering the onset
of a mesa self-replication event that ultimately leads to the creation
of additional mesas. The initiation of this phenomena is studied in
detail for a particular scaling limit of the Brusselator model. Near
the existence threshold $D_c$ of a single steady-state mesa, it is
shown that an internal layer forms in the center of the mesa. The
structure of the solution within this internal layer is shown to be
governed by a certain \emph{core problem}, comprised of a single
non-autonomous second-order ODE. By analyzing this core problem using
rigorous and formal asymptotic methods, and by using the Singular
Limit Eigenvalue Problem (SLEP) method to asymptotically calculate
small eigenvalues, an analytical verification of the conditions of
Nishiura and Ueyema [Physica D, {\bf 130}, No.~1, (1999),
pp.~73--104], believed to be responsible for self-replication, is
given. These conditions include: (1) The existence of a saddle-node
threshold at which the steady-state mesa pattern disappears; (2) the
dimple-shaped eigenfunction at the threshold, believed to be
responsible for the initiation of the replication process; and (3) the
stability of the mesa pattern above the existence threshold. Finally,
we show that the \emph{core problem} is universal in the sense that it
pertains to a class of reaction-diffusion systems, including the
Gierer-Meinhardt model with saturation, where mesa self-replication
also occurs.
\end{abstract}

\label{firstpage}

\setcounter{equation}{0}
\setcounter{section}{0}
\section{{Introduction}}

In \cite{p} Pearson used numerical simulations to show that the
two-component Gray-Scott reaction-diffusion model in the singularly
perturbed limit can exhibit many intricate types of spatially
localized patterns. Many of these numerically computed patterns for
this model have been observed qualitatively in certain chemical
experiments (cf.~\cite{lmps}, \cite{ls}).  An important new phenomenon
that was discovered in \cite{p}, \cite{lmps}, and \cite{ls}, is the
occurrence of self-replication behavior of pulse and spot patterns.
In recent years, many theoretical and numerical studies have been made
in both one and two spatial dimensions to analyze self-replication
behavior for the Gray-Scott model in different parameter regimes
(cf.~\cite{rpp}, \cite{rpp1}, \cite{n5}, \cite{n6}, \cite{u},
\cite{mo}, \cite{dkz}, \cite{dgk1}, \cite{kww-split}). In addition to
the Gray-Scott model, many other reaction-diffusion systems have been
found to exhibit self-replication behavior. These include the 
ferrocyanide-iodide-sulfite system (cf.~\cite{ls}), the
Belousov-Zhabotinsky reaction (cf.~\cite{mpm}), the Gierer-Meinhardt
model (cf.~\cite{gm}, \cite{dp}, \cite{kww-stripe}), and the 
Bonhoffer van-der-Pol-type system (cf.~\cite{ho1}, \cite{ho2}).

 Despite a large number of studies on the subject, the detailed
mechanisms responsible for self-replication are still not clear. In an
effort to classify reaction-diffusion systems that can exhibit pulse
self-replication, Nishiura and Ueyema \cite{n5} (see also \cite{enu})
proposed a set of necessary conditions for this phenomenon to
occur. Roughly stated, these conditions are the following:

\begin{enumerate}
\item{The disappearance of the $K$-spike steady-state solution due to
a saddle-node (or fold point) bifurcation that occurs when a control
parameter is decreased below a certain threshold value.}

\item{The existence of a dimple eigenfunction at the
existence threshold, which is believed to be responsible for the
initiation of the pulse-splitting process. By definition, a dimple
eigenfunction is an even eigenfunction $\Phi(y)$ associated with a
zero eigenvalue, that decays as $|y|\to\infty$ and that has precisely
one positive zero.}

\item{Stability of the steady-state solution above the threshold value for
existence. }

\item{The alignment of the existence thresholds, so that the
disappearance of $K$ pulses, with $K=1,2,3,\ldots$, occurs at
asymptotically the same value of the control parameter. }

\end{enumerate}

 For the Gray-Scott model in the weak interaction parameter regime
where the ratio of the diffusivities is $O(1)$, Nishiura and Ueyema in
\cite{n5} verified these conditions numerically for a given fixed
diffusivity ratio. Alternatively, for the Gray-Scott model in the
semi-strong regime, where the ratio of the diffusivities is
asymptotically large, it was shown in \cite{mo} and in equation (2.9)
of \cite{dkz} that the following core problem determines the spatial
profile of a pulse in the self-replicating parameter regime:
\begin{equation}
V^{\prime\prime}-V+UV^{2}=0\,, \qquad U^{\prime\prime}-UV^{2}=0 \,;
\qquad U^{\prime}\left(0\right)= V^{\prime}\left(  0\right)=0 \,, \quad
  V \rightarrow0 \,, \quad U^{\prime}\rightarrow A \,\,\, \mbox{as} \,\,\,
 y\rightarrow \infty \,. \label{gs:core}
\end{equation}
By using a combination of asymptotic and numerical methods, and
by coupling (\ref{gs:core}) to an appropriate outer solution away from
a localized pulse, conditions (1)--(4) of Nishiura and Ueyema were
verified in \cite{kww-split}. In \cite{dkp} a detailed study of the
intricate bifurcation structure of (\ref{gs:core}) was given.

{ \begin{figure}[tb]
{ \begin{minipage}[t]{0.48\textwidth}
\begin{center}{\small
\setlength{\unitlength}{1\textwidth} \begin{picture}(1,1)(0,0)
\put(-0.1,0.03){\includegraphics[width=1.2\textwidth, height=0.9\textwidth]{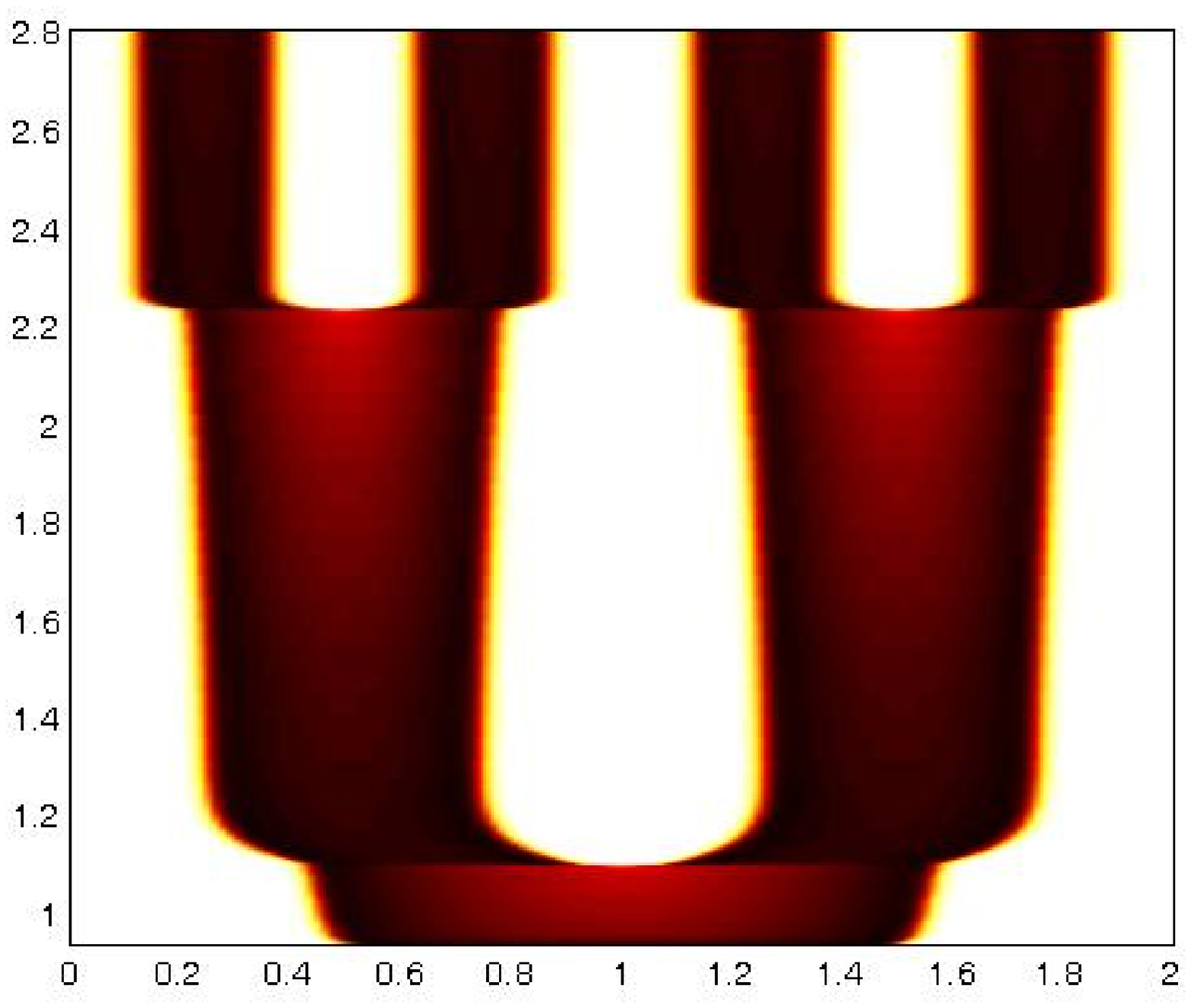}}
\put(-0.1,0.5){$\sqrt{\frac{D_c}{D}}$}
\put(0.5,0.03){$x$}
\end{picture}
}(a)\end{center}
\end{minipage}
\begin{minipage}[t]{0.48\textwidth}
\begin{center}{
\setlength{\unitlength}{1\textwidth} \begin{picture}(1,1)(0,0)
\put(-0.1,0.03){\includegraphics[width=1.2\textwidth, height=0.9\textwidth]{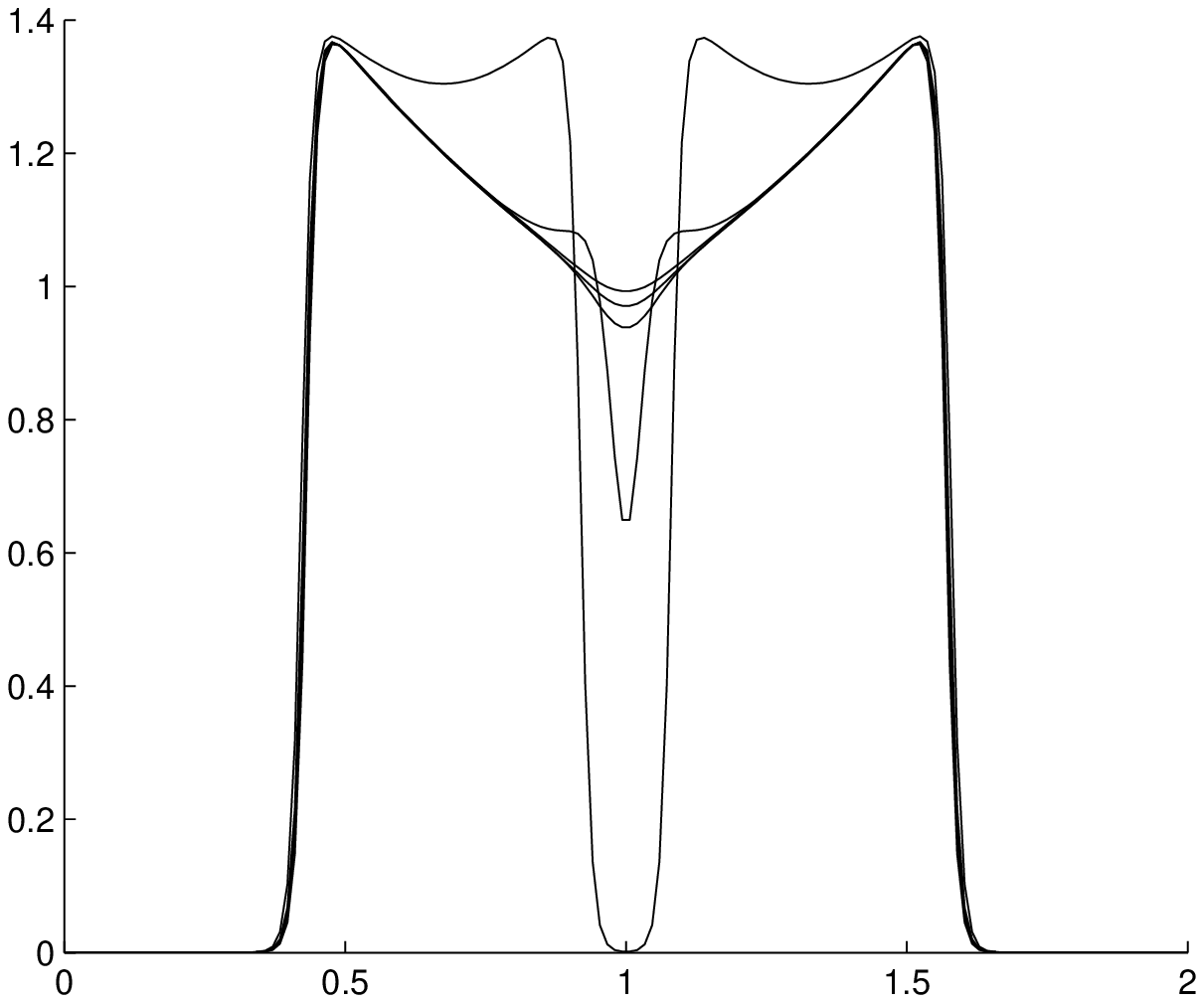}}
\put(0.5,0.03){$x$}
\put(0.1,0.5){$u$}
\end{picture}
}
(b)
\end{center}
\end{minipage}
}\caption{(a) Numerical simulation showing mesa-splitting in the
Brusselator model (\ref{bruv}). The fixed parameters are $\beta_{0} = 1.5$, 
$\eps=0.01$, $\tau=0.7$, $x\in[0,2]$. 
The parameter $D$ is slowly decreased
in time according to the formula $D=(1+5\times10^{-6} t)^{-2}$. 
Parameters $\beta$ and $\alpha$ are
determined through (\ref{30nov2:36}), i.e. $\alpha=\eps^2/D$, $\beta=\alpha
\beta_0.$
The
scale on the vertical axis is $K=\sqrt{\frac{D_{c}}{D}}$, where
$D_{c}=0.88.$ Splitting events occur for $K \approx1$ and $K
\approx2.$ (b) Snapshots of the profile of $u$ during a splitting
event. The time between two successive snapshots is $1000$ time
units. }%
\label{fig:split}%
\end{figure}}

 In this paper we study self-replication of \emph{mesa patterns}. A
single mesa solution is a spatial pattern that consists of two
back-to-back transition layers.  An example of such a steady-state
pattern is shown in Fig.~\ref{fig:ss} below. Our goal is to
analytically verify whether the conditions (1)--(4) of
Nishiura and Ueyama \cite{n5}, originally formulated for analyzing pulse
self-replication behavior, also hold for mesa self-replication. In
addition, we seek to derive and study a certain core problem,
analogous to (\ref{gs:core}), that pertains to self-replicating mesa
patterns. 

For concreteness, we concentrate on the Brusselator model. This model
was introduced in \cite{pl}, and is based on the following
hypothetical chemical reaction:
\[
A\rightarrow X\,, \qquad C+X\rightarrow Y+F \,, \qquad
 2X+Y\rightarrow3X, \qquad X\rightarrow E.
\]
The autocatalytic step $2X+Y\rightarrow3X$ introduces a cubic
non-linearity in the rate equations. Since the 1970's, various
weakly-nonlinear Turing patterns in the Brusselator have been studied
both numerically and analytically in one, two, and three
dimensions. These include spots, stripes, labyrinths and hexagonal
patterns (cf.~\cite{erneux}, \cite{np}, \cite{pena}, \cite{deWit},
\cite{deWit3D}), and oscillatory instabilities and spatio-temporal
chaos (cf.~\cite{k}, \cite{deWitChaos}).

After a suitable rescaling, we write the one-dimensional Brusselator model
on a domain of length $2L$ as
\begin{equation}
u_{t}    =\eps^{2}u_{xx}-u+\alpha+u^{2}v \,, \qquad \tau v_{t}  
  =\eps^{2}v_{xx}+\left(  1-\beta\right)  u-u^{2}v \,;
 \qquad u_{x}(\pm L,t)=v_{x}(\pm L,t)=0 \,.\label{bruv}
\end{equation}
In this paper we make the following assumptions on the parameters:
\begin{equation}
\eps\ll 1\,; \quad \alpha\ll 1\,; \quad \beta\ll 1\,; \quad
D=\frac{\eps^{2}}{\alpha}=O(1) \,; \quad \beta_{0}\equiv\frac{\beta}{\alpha}
 = O(1)\,, \,\,\,\, \mbox{with} \,\,\,\, \beta_0>1\,; \quad
\tau =0 \,. \label{30nov2:36}
\end{equation}
The full numerical results in Fig.~\ref{fig:split} illustrate the
mesa self-replication behavior for (\ref{bruv}). To trigger mesa
self-replication events we started with a single mesa as initial
condition and slowly decreased $D$ in time (see the figure caption for
the parameter values). At the critical value $D_{1}\sim0.8$, a mesa
splits into two mesas, which then repel and move away from each
other. The splitting process is repeated when $D$ is decreased below
$D_{2}~\sim0.2.$

In \S 2 we calculate a threshold value $D_c$ of $D$ for the existence
of a single-mesa steady-state solution for the Brusselator
(\ref{bruv}) in the limit $\eps\to 0$, and under the assumptions
(\ref{30nov2:36}) on the parameter values. The result, summarized in
Proposition \ref{prop:Dc} of \S 2, shows the existence of a value
$D_c$ such that a $K$-mesa steady-state solution exists if and only if
$D>{D_c/K^2}$. Analytical upper and lower bounds for $D_c$ are also
derived. Similar thresholds for the existence of steady-state mesa
patterns were derived for other reaction-diffusion systems in
\cite{ko} using more heuristic means. Our analysis is based on a
systematic use of the method of matched asymptotic expansions.

For a single-mesa steady-state solution, we show in \S 2.1 that an
internal layer of width $O(\eps^{2/3})$ forms in the center of the
mesa when $D$ is asymptotically close to the threshold value
$D_c$. This internal layer is illustrated below in
Fig.~\ref{fig:single-mesa}. By analyzing this internal layer region
using matched asymptotic analysis, we show that the solution $u$ is
determined locally in terms of the solution $U(y)$ to a single
non-homogeneous ODE of the form
\begin{equation}
U^{\prime\prime}=U^{2}-A-y^{2}\,; \qquad \ U^{\prime}(0)=0 \,, \quad
 U^{\prime}(y) \rightarrow1~\ \text{{as\ }}~y\rightarrow\infty
 \,. \label{intro:core}%
\end{equation}
Here $A$ is related to the parameter values in (\ref{bruv}).  We refer
to (\ref{intro:core}) as the \emph{core problem} for the onset of
self-replication. Unlike (\ref{gs:core}) for self-replicating pulses
in the Gray-Scott model, the problem (\ref{intro:core}) is not coupled
and, consequently, is easier to study analytically than
(\ref{gs:core}). The proof of conditions (1) and (2) of
Nishiura and Ueyama is then reduced to a careful study of
(\ref{intro:core}). More specifically, by using rigorous techniques we
prove analytically the existence of a saddle-node bifurcation for
(\ref{intro:core}) and we analyze the solution behavior on the
bifurcation diagram. The result is summarized below in Theorem
\ref{thm:core}. In \S 2.2 we use some rigorous properties of the core
problem, together with a formal matched asymptotic analysis, to
asymptotically construct a dimple eigenfunction corresponding to the
zero eigenvalue at the saddle-node bifurcation value. This
construction, summarized in Proposition \ref{thm:cond2}, establishes
condition (2) of Nishiura and Ueyama.

In \S 2.3 we show that the core problem (\ref{intro:core}) is
universal in the sense that it can be readily derived for other
reaction-diffusion systems where mesa self-replication occurs. The
universal nature of (\ref{intro:core}) is illustrated for some
specific systems, including the Gierer-Meinhardt model with saturation
(cf.~\cite{gm}). For this specific model, mesa-splitting was computed
numerically in Figure 28 of \cite{kww-stripe}. Although the phenomena
of mesa self-replication is qualitatively described in Chapter 11 of
\cite{ko},  the core problem
(\ref{intro:core}) governing the onset of mesa self-replication and its
analysis has not, to our knowledge, appeared in the literature.

Since the saddle-node existence value for a $K$-mesa steady-state
solution is $D={D_{c}/K^2}$, the condition (4) of Nishiura
and Ueyama, regarding an asymptotically close alignment of saddle-node
bifurcation values, does not hold in a strict sense. However, this
condition is satisfied in the same approximate sense as in the study
of self-replicating pulses for the Gray-Scott model in the semi-strong
interaction regime (see Table 3 of \cite{dgk1} and equation (1.2) of
\cite{kww-split}).

In \S \ref{sec:stability} we study the stability of $K$-mesa
steady-state solutions when $D>{D_{c}/K^2}$. We show that such a
pattern is stable when $\tau=0$, and moreover all asymptotically small
eigenvalues are purely real. This proves condition (3) of Nishiura and
Ueyema. Our analysis is similar to the SLEP\ method, originally
developed by Nishiura et. al. in \cite{n1} and \cite{n2}, and that has
been used successfully to prove the stability of mesa-type patterns in
reaction-diffusion systems and in related contexts (cf.~\cite{n3},
\cite{n4}). In our analysis a formal matched asymptotic analysis is
used to derive a reduced problem that capture the asymptotically small
eigenvalues of the linearization. This reduced system is then studied
rigorously using several tools, including the maximum principle and
matrix theory. In this way we prove that the small eigenvalues are
purely real and negative when $\tau=0$.

{ \begin{figure}[tb]
{ \begin{minipage}[t]{0.48\textwidth}
\begin{center}{\small
\setlength{\unitlength}{1\textwidth} \begin{picture}(1,1)(0,0.05)
\put(0.08,0.128){\includegraphics[width=0.9\textwidth, height=0.77\textwidth]{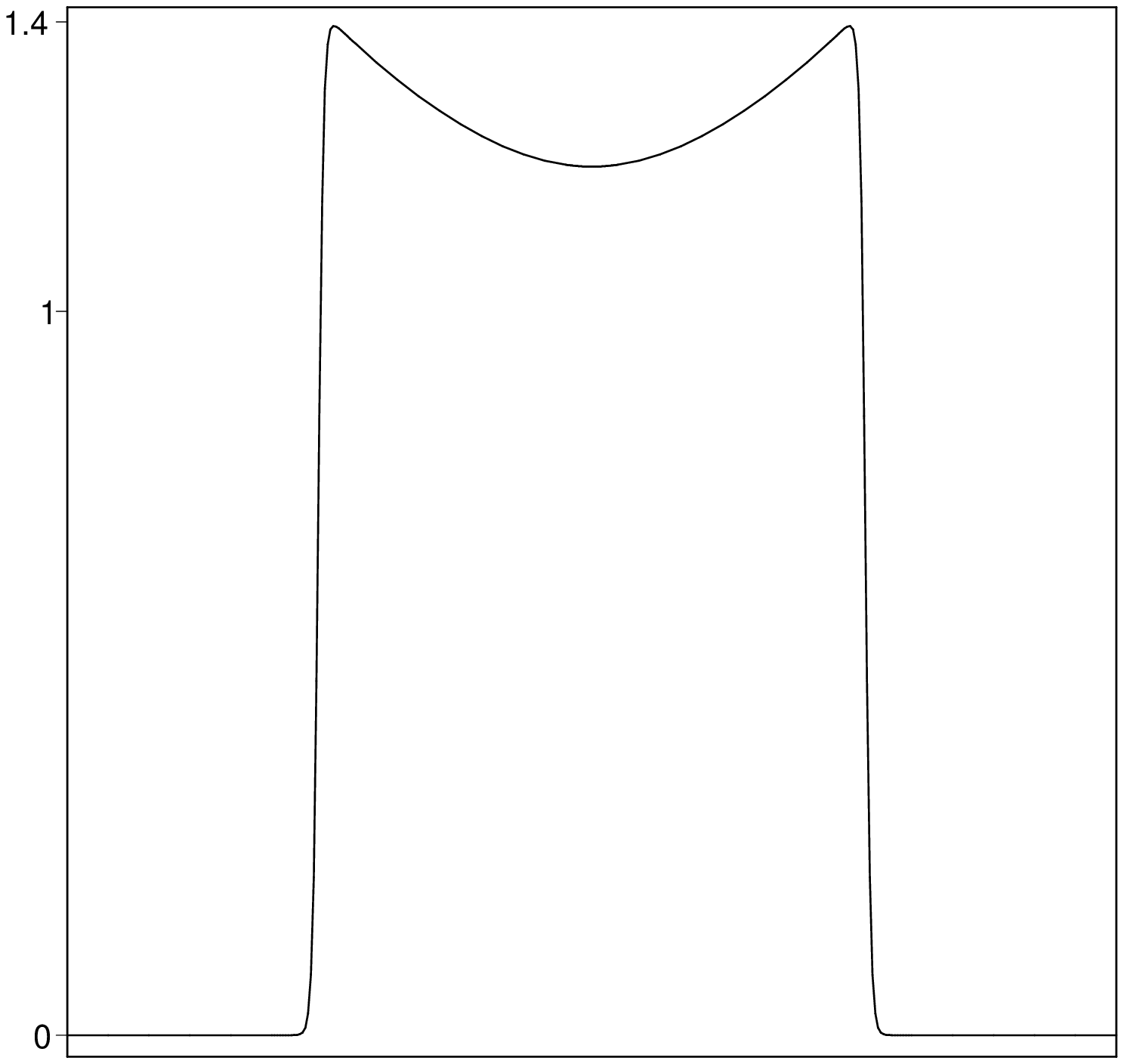}}
\put(0.08,0.04){$-L$}
\put(0.28,0.04){$-l$}
\put(0.75,0.04){$l$}
\put(0.95,0.04){$L$}
\end{picture}
}(a)\end{center}
\end{minipage}
\begin{minipage}[t]{0.48\textwidth}
\begin{center}{
\setlength{\unitlength}{1\textwidth} \begin{picture}(1,1)(0,0.05)
\put(0.08,0.1){\includegraphics[width=0.9\textwidth, height=0.8\textwidth]{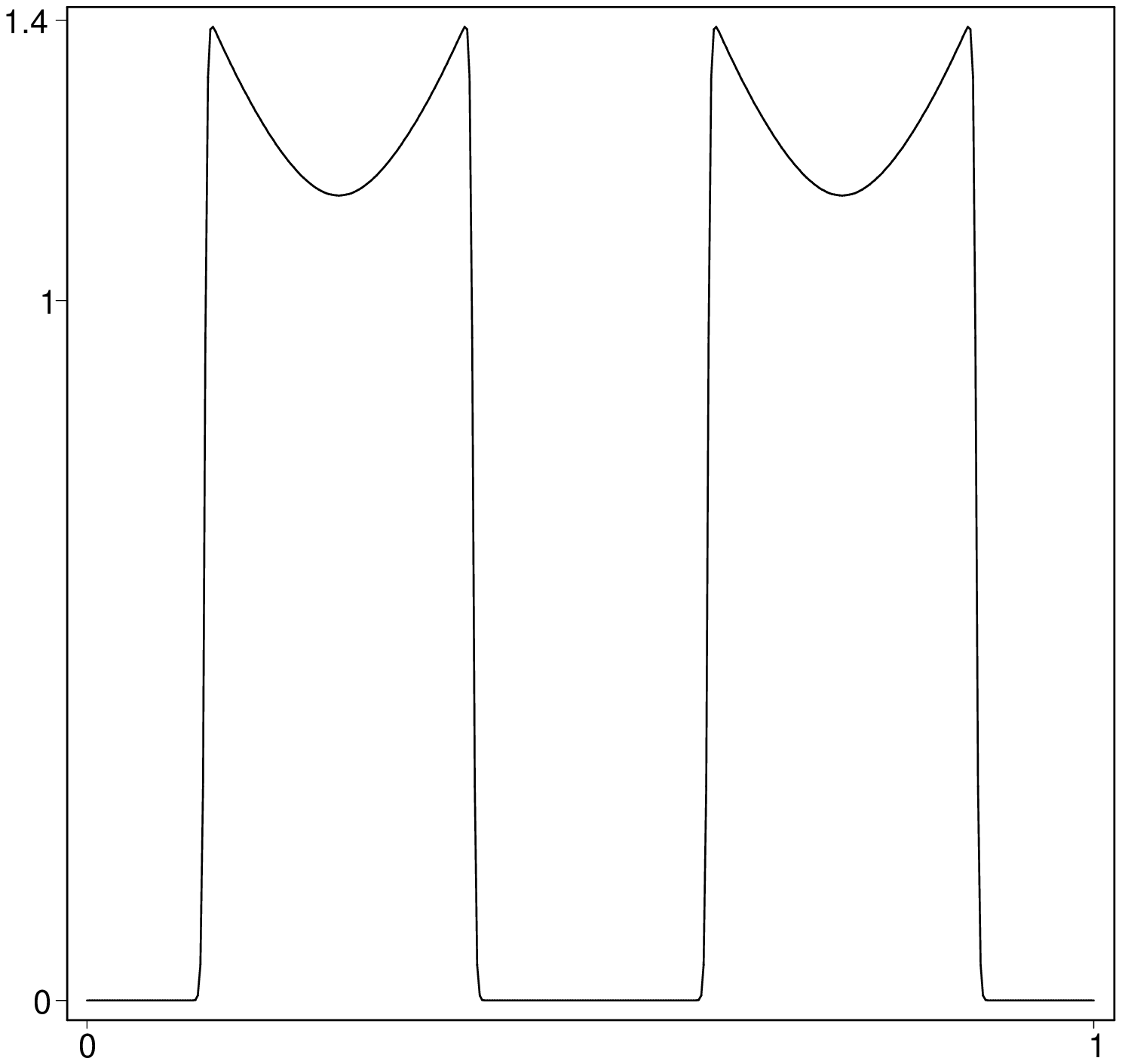}}
\put(0.21,0.08){$x_{1-}$}
\put(0.45,0.08){$x_{1+}$}
\put(0.61,0.08){$x_{2-}$}
\put(0.85,0.08){$x_{2+}$}
\end{picture}
}
(b)
\end{center}
\end{minipage}
}\caption{(a) A single mesa steady-state on $[-L, L]$. The
parameter values are $L=1,~\beta_{0}=1.5,~D=1.29,~\eps=0.005.$ (b)
A two-mesa steady state $(K=2, L=1/4, ~x\in[0,1])$. The parameter values are
$~D=0.068, ~\eps=0.00125, ~\beta_{0} = 1.5.$}%
\label{fig:ss}%
\end{figure}}

Finally, in \S \ref{sec:discuss} we relate our results regarding mesa
self-replication for the Brusselator model with previous results
concerning the coarsening phenomenon of mesa patterns that occurs when
$D$ is sufficiently large (cf.~\cite{kew}). In addition, we propose
some open problems.

\setcounter{equation}{0}
\setcounter{section}{1}
\section{The Steady State Mesa and the Universal Core Problem}\label{sec:split}

 In this section we study the steady-state problem for (\ref{bruv}),
and we prove analytically the first two conditions of Nishiura and
Ueyema.  We will analyze an even symmetric solution of the type shown
on Fig.~\ref{fig:ss}(a), consisting of a single mesa on a domain
$[-L,L]$ with interfaces at $x=\pm l$. The $K$-mesa solution on a
domain of length $2KL$ can then be constructed by reflecting and
gluing together $K$ such solutions.

We first reformulate (\ref{bruv}) to emphasize the slow-fast
structure. We define $w$ by
\begin{equation*}
    w=v+u \,,
\end{equation*}
so that (\ref{bruv}) becomes
\begin{equation}
u_{t}    =\eps^{2}u^{\prime\prime}-u+\alpha+u^{2}(w-u) \,, \qquad
\tau\left(  w_{t}-u_{t}\right)  +u_{t}    =\eps^{2}w^{\prime\prime
}+\alpha-\beta  \,. \label{wsys1}
\end{equation}
Here the primes indicate derivatives with respect to $x$. 
We then introduce $\beta_0=O(1)$ and $D=O(1)$ defined by
\begin{equation*}
\beta_{0}\equiv\frac{\beta}{\alpha}\,, \qquad D\equiv\frac{\eps^{2}}
{\alpha} \,.
\end{equation*}
Then, (\ref{wsys1}) becomes
\begin{equation}
u_{t}    =\eps^{2}u^{\prime\prime}-u+\alpha+u^{2}(w-u) \,, \qquad
\frac{\tau}{\alpha} w_{t}  + \frac{(1-\tau)}{\alpha} u_t
  = D w^{\prime\prime}+1-\beta_0  \,. \label{wsys}
\end{equation}
The corresponding steady-state problem is
\begin{equation}
 \eps^{2}u^{\prime\prime}-u+u^{2}(w-u) + \alpha =0 \,, \qquad
   D w^{\prime\prime}+1-\beta_{0}u =0 \,, \label{ssuw}
\end{equation}
where $\alp=O(\eps^2)$. Since $\eps^{2}\ll D$ from (\ref{30nov2:36}),
it follows that $w$ is the slow variable and $u$ is the fast variable.
Upon integrating (\ref{ssuw}) for $w$ and using the Neumann boundary
condition for $w$ at $x=L$, and the symmetry condition
$w^{\prime}(0)=0$, we obtain the integral constraint
\begin{equation}
L=\beta_{0}\int_{0}^{L} u \, dx \,. \label{u0mass}
\end{equation}

Near the interface at $x=l$ we introduce the inner expansion
\begin{equation}
 u  =U_{0}\left(  y\right)  + \eps U_{1}\left(  y\right)
+\ldots,\ \ \ \ \ w=W_{0}+\eps W_{1}\left(  y\right)  +\ldots \,, \qquad
  y=\eps^{-1}(x-l) \,. \label{eq:inn}
\end{equation}
Upon substituting this expansion into (\ref{ssuw}), we obtain the
leading-order problem
\begin{equation}
  U_{0}^{\prime\prime}-f(U_{0},W_{0})=0 \,, \qquad
 W_{0}^{\prime\prime}=0 \,, \label{U0}
\end{equation}
where $f(u,w)$ is defined by
\begin{equation}
f(u,w)\equiv u-u^{2}(w-u) \,. \label{fdef}
\end{equation}
At next order, we obtain
\begin{equation}
 {\cal L} U_1 \equiv
  U_{1}^{\prime\prime}-f_u(U_{0},W_{0}) U_1 = f_w(U_0,W_0)W_1 \,, \qquad
 W_{1}^{\prime\prime}=0 \,. \label{U1}
\end{equation}

From (\ref{U0}) we get that $W_{0}$ is a constant to be determined. To
ensure that there exists a heteroclinic connection for $U_0$ we
require that $f$ satisfy the \emph{Maxwell line condition}, which states
that the area between the first two roots of $f$ is the negative
of the area between its last two roots of $f.$ Since $f$ is a cubic,
this is equivalent to simultaneously solving $f=0$ and
$f^{\prime\prime}=0$ for $W_{0}$. In this way, we obtain 
\begin{equation}
W_{0}=\frac{3}{\sqrt{2}} \,, \qquad U_{0}=\frac{1}{\sqrt{2}}\left[
  1\pm\tanh\left(  \frac{y}{2}\right) \right] \,. \label{U0W0}
\end{equation}
For the mesa solution as shown in Fig.~\ref{fig:ss}(a), we must take
the minus sign in (\ref{U0W0}) above.

 To determine the interface location $l$, we now study the outer
problem away from the interface at $x=l$. Since $\alp=O(\eps^2)$, we
obtain to leading order from (\ref{ssuw}) that
\begin{equation*}
  u+u^{2}(w-u) = 0 \,.
\end{equation*}
This yields either $u=0$ or 
\begin{equation}
   w\sim h\left(  u\right) \equiv \frac{1}{u}+u \,. \label{w=h(u)}%
\end{equation}
Moreover, we have $U_{0}\rightarrow0$ as $y\rightarrow\infty$ and
$U_{0}\rightarrow\sqrt{2}$ as $y\rightarrow-\infty$. Therefore, by
matching to $U_0$ and $W_0$, and by using the symmetry condition
at $x=0$, we obtain the following outer problem in the mesa region 
$0\leq x\leq l$:
\bsub \label{outer}
\begin{gather}
 w=h(u); \qquad Dw^{\prime\prime}=g\left(  u\right)  \equiv\beta
_{0}u-1 \,, \qquad 0<x<l \,.\label{outer1} \\
 u\left(  l\right)  =\sqrt{2}\,, \qquad w\left(  l\right)  =\frac{3}{\sqrt{2}}
\,, \qquad u^{\prime}\left(  0\right) =w^{\prime}\left(  0\right)=0 \,.
   \label{outer2}
\end{gather}
\esub
The leading-order outer problem on $l\leq x\leq L$ is $u=0$ and 
$Dw^{\prime\prime}=-1$.

The solution to the second-order inner problem (\ref{U1}) for $W_1$ is
$W_{1}=W_{11}y+ W_{12}$, where $W_{11}$ and $W_{12}$ are constants to
be determined. Since ${\cal L} U_{0}^{\prime}=0$, the solvability condition
for (\ref{U1}) yields
\begin{equation*}
 0= \int_{-\infty}^{\infty} U_{0}^{\prime} f_{w}(U_0,W_0) W_1 \, dy
  = -\int_{-\infty}^{\infty} U_{0}^{\prime} U_{0}^2 \left(W_{11} y + W_{12}
 \right) \, dy \,.
\end{equation*}
This yields one relation between $W_{11}$ and $W_{12}$. The second relation
is obtained by matching $W$ to the the outer solution $w$. This yields
$W_{11}=w^{\prime}(l^{\pm})$. In this way, we obtain
\begin{equation}
   W_{1}^{\prime}\equiv W_{11}=w^{\prime}(l) \,, \qquad
  W_{12}=-\frac{W_{11}}{2\sqrt{2}} \int_{-\infty}^{\infty} y \left(U_{0}^3
  \right)^{\prime} \, dy \,. \label{w1prime}
\end{equation}

We now solve the outer problem (\ref{outer}) in terms of $u_0\equiv
u(0)$. We first define $F\left(u;u_0\right)$ by
\begin{equation}
F\left(  u;u_{0}\right)  \equiv\int_{u_{0}}^{u}g\left(  s\right)  h^{\prime
}(s) \, ds \,. \label{fcap}
\end{equation}
By multiplying (\ref{outer1}) for $w$ by $w^{\prime}$ we get
\begin{equation*}
D\frac{w^{\prime2}}{2}=F\left(  u;u_0\right) \,, \qquad
 w^{\prime} = \sqrt{ \frac{2}{D}} \sqrt{ F(u;u_0)} \,.
\end{equation*}
In the outer region on $l\le x\le L$, we have $u=0$. Therefore, by
integrating $w^{\prime\prime}$ from $x=0$ to $x=L$, we obtain
$\int_{0}^{l} g(u)\, dx +\int_{l}^{L} (-1) \, dx=0$. This yields,
\begin{equation}
\int_{0}^{l} g(u) \, dx = L - l \,. \label{const}
\end{equation}
The left-hand side of (\ref{const}) is evaluated by
integrating $w^{\prime\prime}$ from $x=0$ to $x=l$ to
get $\int_{0}^{l}g\left(  u\right)  dx=Dw^{\prime}\left(  l\right)
=\sqrt{(2 D) F(\sqrt{2};u_{0})}$. In addition, by using
$w^{\prime}=h^{\prime}(u)u^{\prime}$, we obtain
\begin{equation}
\frac{du}{dx}= \sqrt{\frac{2F\left(  u;u_0\right)  }{D}} \left[h^{\prime}(u)
 \right]^{-1} \,. \label{uone}
\end{equation}
We then integrate (\ref{uone}) with $u(0)=u_0$ and $u(l)=\sqrt{2}$. In
this way, we obtain
\begin{equation}
 \sqrt{2 F(\sqrt{2};u_{0})}= \frac{L-l}{\sqrt{D}} \,, 
 \qquad \frac{l}{\sqrt{D}} =\int_{u_{0}}^{\sqrt{2}}%
\frac{h^{\prime}\left(  u\right)  }{\sqrt{2F\left(  u;u_0\right)  }}
 \, du \,. \label{temp1}
\end{equation}
Upon integrating the second expression in (\ref{temp1}) by parts we get
\[
\ \int_{u_{0}}^{\sqrt{2}}\frac{h^{\prime}\left(  u\right)  }{\sqrt{2F\left(
u;u_0\right)  }}\, du =\frac{\sqrt{2F\left(  \sqrt{2};u_{0}\right)  }}
 {g\left(  \sqrt
{2}\right)  }+\int_{u_{0}}^{\sqrt{2}}\frac{g^{\prime}\left(  u\right)  }%
{ \left[g\left(  u\right)\right]^2}\sqrt{2F\left(  u;u_0\right) } \, du \,.%
\]
By combining this relation with (\ref{temp1}), and by calculating
$g(\sqrt{2})$, we obtain the following expression relating $u_0$ to
the overall length of the domain:
\begin{equation}
\chi\left(  u_{0}\right)  \equiv\sqrt{2F(\sqrt{2};u_{0})}\left(  \frac
{\beta_{0}\sqrt{2}}{\beta_{0}\sqrt{2}-1}\right)  +\int_{u_{0}}^{\sqrt{2}}%
\frac{g^{\prime}\left(  u\right)  }{ \left[g\left(  u\right)\right]^2 }
 \sqrt{2F\left(u;u_{0}\right)  } \, du= \frac{L}{\sqrt{D}} \,. \label{chidef}
\end{equation}

We note that the function $h\left( u\right) $ has a minimum at
$u_{0}=1$, and that $\frac{dF}{du_{0}}=-g\left( u_{0}\right)
h^{\prime}\left( u_{0}\right) <0$ for $u_{0}>1$ from
(\ref{fcap}). Hence, $\chi\left( u_{0}\right) $ is a decreasing
function of $u_0$ for $u_{0}>1$.  Upon defining $D_c$ by
$D_{c}={L^{2}/\chi\left( 1\right)^{2}}$, we conclude that there exists
a value of $u_0$, and consequently an outer solution on $0<x<l$
exists, if and only if $D\ge D_c$. Therefore, we obtain a threshold
value $D_c$ for the existence of a single mesa solution on $[-L,L]$.

Next, we obtain explicit bounds on $D_{c}$. The function
$w=h\left( u\right) $ satisfies $h\left( 1\right) =2$ and $h\left( \sqrt
{2}\right) =3/\sqrt{2}$. Therefore, for  $x\in\lbrack0,l]$ we have
that $u$ and $w$ are both increasing functions for
$x\in\lbrack0,l]$ with $u\left(  l\right) =\sqrt{2}$,
$w\left(  l\right)  = {3/\sqrt{2}}$, and
\begin{equation}
1<u<\sqrt{2} \,, \qquad  2<w<3/\sqrt{2} \,.\label{8:24apr10}
\end{equation}
Using ($\ref{8:24apr10}$) and (\ref{outer1}) for $w$ we estimate
$w^{\prime\prime} = D^{-1}\left(  \beta_{0}u-1\right) \geq D^{-1}
(\beta_{0}-1)$. This yields that
\begin{equation}
w\left(  x\right)    \geq w\left(  0\right)  +\frac{(\beta_{0}-1)}{2D}x^{2}
  \,, \qquad \mbox{which implies} \qquad
\frac{3}{\sqrt{2}}   \geq2+\frac{(\beta_{0}-1)}{2D}l^{2} \,. \label{dbound}
\end{equation}
Then, we use (\ref{u0mass}) with $u=0$ for $l\leq x\leq L$ to get
\begin{equation*}
L=\beta_{0}\int_{0}^{l}u \, dx \leq \beta_{0}l\sqrt{2} \,.
\end{equation*}
Combining this inequality with (\ref{dbound}), we get
\begin{equation*}
  l^{2}\geq\frac{L^{2}}{2\beta_{0}^{2}} \,, \qquad
\frac{3}{\sqrt{2}}\geq2+\frac{\beta_{0}-1}{4\beta_{0}^{2}}\frac{L^{2}}{D}.
\end{equation*}
By solving this second relation for $D$ we conclude that a necessary
condition for the existence of a solution to (\ref{outer}) when
$\eps\ll 1$ is that $D\geq\underline{D}$, where
\begin{equation}
\underline{D}=\frac{(\beta_{0}-1)}{4\beta_{0}^{2}}\frac{1}{3/\sqrt{2}-2}L^{2}
 \,.
\end{equation}
This relation provides a lower bound for $D_{c}$. To find an upper bound we
calculate $w^{\prime\prime}\leq{\left(  \beta_{0}\sqrt{2}-1\right)/D}$.
This inequality can be integrated, and with $w(l)={\sqrt{3}/2}$, we get 
\begin{equation*}
 w\left(  0\right) \geq\frac{3}{\sqrt{2}}-\frac{l^{2}}{2D}
 \left(\beta_{0}\sqrt{2}-1\right) \,.
\end{equation*}
Then, from (\ref{u0mass}), we obtain
\begin{equation}
l\leq\frac{L}{\beta_{0}} \,.
\end{equation}
The condition $w\left(  0\right)  \geq2$ is satisfied provided that $\frac
{3}{\sqrt{2}}-\frac{L^{2}}{2D}\frac{1}{\beta_{0}^{2}}\left(  \beta_{0}\sqrt
{2}-1\right)  >2.$ Hence, a sufficient condition for the existence of a
solution to (\ref{outer}) when $\eps\ll 1$ is that 
$D\geq\overline{D}$, where
\begin{equation*}
\overline{D}=\frac{\sqrt{2}\beta_{0}-1}{2\beta_{0}^{2}}\frac{1}{3/\sqrt{2}
-2}L^{2} \,.
\end{equation*}
It follows that $\underline{D}\leq D_{c}\leq\overline{D}$. We summarize the
results of this analysis of a single mesa solution as follows:

{ \begin{figure}[tb]
\begin{center}
{ { \setlength{\unitlength}{1\textwidth} \begin{picture}(1,0.5)(0,0)
\put(0.03,0.02){\includegraphics[width=0.9\textwidth]{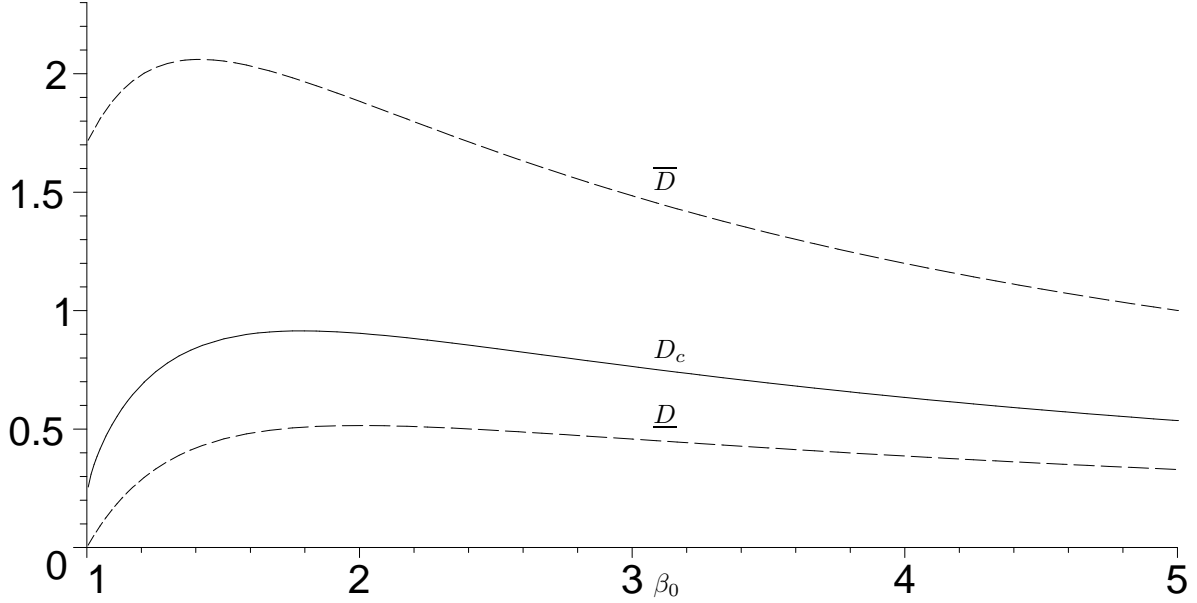}}
\put(0.52,0.02){$\beta_0$}
\put(0.52,0.33){$\overline{D}$}
\put(0.52,0.2){$D_c$}
\put(0.52,0.15){$\underline{D}$}
\end{picture}
} }
\end{center}
\caption{ {The graphs of $D_{c}$, }$\overline{D}${\ and }$\underline{D}${ }
{versus $\beta_{0}$ } }%
\label{fig:beta0Dc}%
\end{figure}}

\begin{proposition}
[Nishiura-Ueyema's Condition 1: The Steady State and its
Disappearance]{\label{prop:Dc} Consider the steady state solution to the 
Brusselator (\ref{ssuw}) with $\beta_{0}>1$ in the limit $\eps\goesto 0$. 
We define $F(u;u_0)$ and $\chi(u_0)$ by
\begin{equation*}
F\left(  u;u_{0}\right)  \equiv\int_{u_{0}}^{u}\left(  \beta_{0}s-1\right)
\left(  1-\frac{1}{s^{2}}\right)  ds \,,  \qquad
\chi\left(  u_{0}\right)  \equiv\sqrt{2F(\sqrt{2};u_{0})}\left(  \frac
{\beta_{0}\sqrt{2}}{\beta_{0}\sqrt{2}-1}\right)  +\int_{u_{0}}^{\sqrt{2}}%
\frac{\beta_{0} \sqrt{2F\left(u;u_{0}\right)  }}
 {\left(  \beta_{0}u-1\right)^{2}}  \, du \,.
\end{equation*}
In terms of $\chi(1)$, we define the threshold $D_c$ by
\begin{equation*}
D_{c}= {L^{2}/\chi\left(  1\right)  ^{2}} \,.
\end{equation*}
Suppose that $D>D_{c}.$ Then, there exists a $u_{0}\in\left(  1,\sqrt{2}\right)
$ and $l\in\left(  0,L\right)  $ given implicitly by
\begin{equation*}
\chi\left(  u_{0}\right)  =\frac{L}{\sqrt{D}}, \qquad l=L-\sqrt{D}\sqrt
{2F(\sqrt{2};u_{0})} \,, 
\end{equation*}
such that there exists a symmetric mesa solution on the interval $[-L,L]$ 
with interfaces at $\pm l$ and with $u\left(  0\right)  =u_{0}.$ In the 
region $x\in\left(  0,l\right)$, $w$ and $u$ are given implicitly by
\begin{equation}
w   =\frac{1}{u}+u\,, \qquad  Dw^{\prime\prime}=g\left(  u\right)  \equiv
\beta_{0}u-1 \,, \quad  0<x<l \,; \qquad u\left(  0\right) =u_{0} \,, 
 \quad u\left(  l\right)  =\sqrt{2} \,. \label{eqw}
\end{equation}
In the region $x\in\left(  l,L\right)$, the leading-order outer solutions 
for $u$ and $w$ are
\begin{equation*}
u=0 \,, \qquad w\sim-\frac{1}{2D}\left(  x-L\right)  ^{2}+\frac{1}{2D}\left(
2/\sqrt{3}-L\right)  ^{2}+2/\sqrt{3} \,.
\end{equation*}
Moreover, we have $\underline{D}\leq D_{c}\leq\overline{D}$ where
\begin{equation*}
\underline{D}=\frac{(\beta_{0}-1)}{4\beta_{0}^{2}}\frac{1}{3/\sqrt{2}-2}L^{2}
 \,, \qquad
\overline{D}=\frac{\sqrt{2}\beta_{0}-1}{2\beta_{0}^{2}}\frac{1}{3/\sqrt{2}%
-2}L^{2} \,.
\end{equation*}
The graphs of $D_{c}$, $\overline{D}$, and $\underline{D}$, versus 
$\beta_{0}$ are shown in Fig.~\ref{fig:beta0Dc}. A single-mesa solution 
does not exist if $D<D_{c}$. By reflections and translations, a single-mesa 
solution can be extended to a $K$-mesa solution on an interval of length 
$2KL$.}
\end{proposition}

\subsection{{Core problem\label{sec:core}}}

In this section we verify Nishiura and Ueyema's second condition by
analyzing a limiting differential equation that is valid in the
vicinity of the critical threshold $D=D_{c}.$ When $D$ is decreased
slightly below $D_{c}$ (at which point $u(0)\sim1,w(0)\sim2)$ the
single-mesa solution ceases to exist. To study the solution near this
fold point, we fix $u(0)=u_{0}$ and consider $D=D(u_{0})$. From
numerical computations of the steady state solution as shown in
 Fig.~\ref{fig:single-mesa}, an internal layer forms near the origin when
$u_{0}$ is decreased below $u(0)=1$.

{ \begin{figure}[tb]
\begin{center}
{ { \setlength{\unitlength}{1\textwidth} \begin{picture}(1,0.5)(0,0)
\put(0.03,0.02){\includegraphics[width=0.9\textwidth]{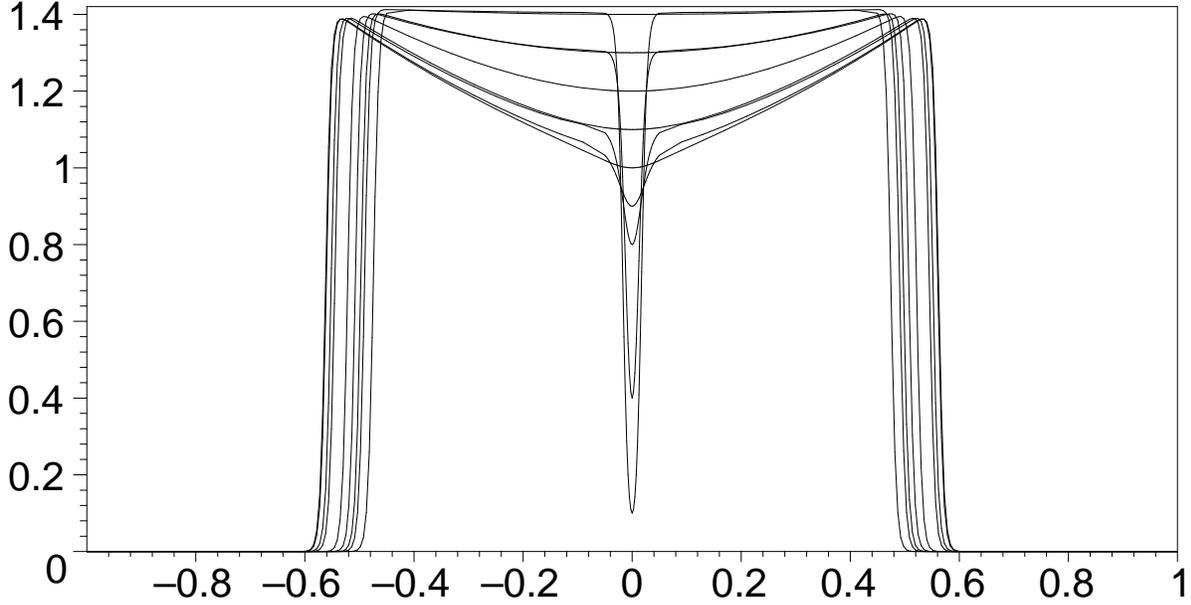}}
\end{picture}
} }
\end{center}
\caption{ Formation of a boundary layer near the center of a mesa. The steady
state and $D$ are solved simultaneously, while $u(0)$ is fixed at one of the
values $1.4,1.3,1.2,1.0,0.9,0.4,0.1.$ The corresponding values for $D$ are as
follows: $u(0)=1.4, D=17.109; ~~ u(0)=1.3, D=2.257; ~~ u(0)=1.2, D=1.290; ~~
u(0)=1.1, D=0.96; ~~ u(0)=1.0, D=0.846; ~~ u(0)=.9, D=0.863; ~~ u(0)=.8,
D=0.938; ~~ u(0)=.7, D=1.068; ~~ u(0)=.6, D=1.27; ~~ u(0)=.5, D=1.624; ~~
u(0)=.4, D=2.244; ~~ u(0)=.3, D=3.525; ~~ u(0)=.2, D=6.982; ~~ u(0)=.1,
D=24.34. $ }%
\label{fig:single-mesa}%
\end{figure}}

To study the initial formation of this internal layer near the origin,
we expand (\ref{ssuw}) near $D=D_{c}$ as
\begin{equation*}
u =1+\delta u_{1}+\cdots \,, \qquad w   =2+\delta^{2}w_{1}+\cdots \,, 
 \qquad D   =D_{c}+\cdots \,.
\end{equation*}
The nonlinear term in (\ref{ssuw}) becomes 
$f(u,w)=\delta^2(u_1^2-w_1)+\cdots$. From (\ref{ssuw}) we then obtain
\begin{equation}
\frac{\eps^{2}}{\delta}u_{1}^{\prime\prime}    =u_{1}^{2}-w_{1} \,, \qquad
D_{c}\delta^{2}w_{1}^{\prime\prime}   =\beta_{0}-1 \,. \label{core:temp1}
\end{equation}
We introduce the inner-layer variable $z$ by $z={x/\delta}$ with $\delta\ll
1$. Then, (\ref{core:temp1}) for $w_1$ becomes
\begin{equation*}
 w_{1zz}    =\frac{\beta_{0}-1}{D_{c}} \,.
\end{equation*}
The solution is
\begin{equation}
w_{1}   =\mathcal{A}+\mathcal{B}z^{2} \,, \qquad \mathcal{B}=\frac
{\beta_{0}-1}{2 D_{c}}>0 \,, \qquad \mathcal{A}=w_{1}\left(  0\right)\,.
  \label{calB}
\end{equation}
Then, (\ref{core:temp1}) for $u_1$ is
\begin{equation}
\frac{\eps^{2}}{\delta^{3}}u_{1zz}=u_{1}^{2}-\left(  \mathcal{A}%
+\mathcal{B}z^{2}\right)  \,. \label{22nov3:53}%
\end{equation}
This suggests the internal-layer scaling $\delta=\eps^{2/3}$ so that
$u_{1zz}=u_{1}^{2}-\left(  \mathcal{A}+\mathcal{B}z^{2}\right)$. The
boundary conditions for $u_1$ are
$u_{1z}\left(  0\right)=0$ and $u_{1}\rightarrow
z\sqrt{\mathcal{B}}$ as $z\rightarrow\infty$. Finally, we introduce
$U$ and $y$ as 
\begin{equation*}
u_{1}=\mathcal{B}^{1/3}U \,, \qquad  z=\mathcal{B}^{-1/6}y\,.
\end{equation*}
This yields the following {\em core problem} for $U(y)$ on $0<y<\infty$:
\begin{equation}
U^{\prime\prime}    =U^{2}-A-y^{2} \,; \qquad
U^{\prime}(0)   =0 \,,  \qquad U^{\prime} \sim 1 \quad \mbox{as} \quad
 y\rightarrow \infty \,.\label{core}\\
\end{equation}
Here $A$ is related to $\mathcal{A}$ and $\mathcal{B}$ by
\[
A=\mathcal{A\mathcal{B}}^{-2/3} \,.
\]
In terms of the original variables, we have that
\bsub \label{trans}
\begin{gather}
     u(x) - 1 \sim \eps^{2/3} \mathcal{B}^{1/3} U(y) \,, \qquad
     w(x) -2  \sim \eps^{4/3} \mathcal{B}^{2/3} \left(A + y^2 \right) \qquad
    y = x/(\eps^{2/3} \mathcal{B}^{-1/6}) \,, \label{trans1} \\
     u_{1}(0)={\mathcal B}^{1/3} U(0) =\eps^{-2/3}\left[u(0)-1\right]\,, \qquad
     w_{1}(0)={\mathcal B}^{2/3} A =\eps^{-4/3}\left[w(0)-2\right] \,.
  \label{trans2}
\end{gather}
\esub
Here $\mathcal{B}$ is defined in (\ref{calB}).  The following main
result pertains to the solution behavior of (\ref{core}).

\begin{theorem}
{\label{thm:core} Suppose that $A\gg1.$ Then, the core problem
(\ref{core})\ admits exactly two solutions $U^{\pm}(y)$ with
$U^{\prime}>0$ for $y>0$. They have the following uniform expansions:
\bsub \label{U+-}
\begin{align}
U^{+} &\sim\sqrt{A+y^{2}} \,, \qquad  U^{+}\left(  0\right)  \sim\sqrt{A}
\,,  \label{U+} \\
 U^{-}&\sim\sqrt{A+y^{2}}\left(  1-3\ \operatorname{sech}{}^{2}\left(
\frac{A^{1/2}y}{\sqrt{2}}\right)  \right)  \,, \qquad
U^{-}\left(  0\right)  \sim-2\sqrt{A} \,. \label{U-}
\end{align}
\esub
These two solutions are connected. For any such solution, let
$s=U(0)\equiv U_{0}$ and consider the solution branch $A=A\left( s\right)$. 
Then, $A(s)$ has a unique (minimum) critical point at $s=s_{c},\ $ $A=A_{c}.$
Moreover, define $\Phi(y)$ by
\begin{equation}
\Phi=\frac{\partial U}{\partial s}\Big{|}_{s=s_{0}} \,. \label{phi0}
\end{equation}
Then, $\Phi>0$ for all $y\geq0$ and $\Phi\rightarrow0$ as $y\rightarrow0.$
Numerically, we calculate that $A_{c}\approx-1.46638$ and
$s_{c}\approx-0.61512.$ The graph of $A\left(  s\right)  $ is shown in
Fig.~\ref{fig:sA}.}
\end{theorem}

{ \begin{figure}[tb]
\begin{center}
{\small \setlength{\unitlength}{1\textwidth} \begin{picture}(1,0.6)(0,0)
\put(0.08,0.1){\includegraphics[width=0.9\textwidth]{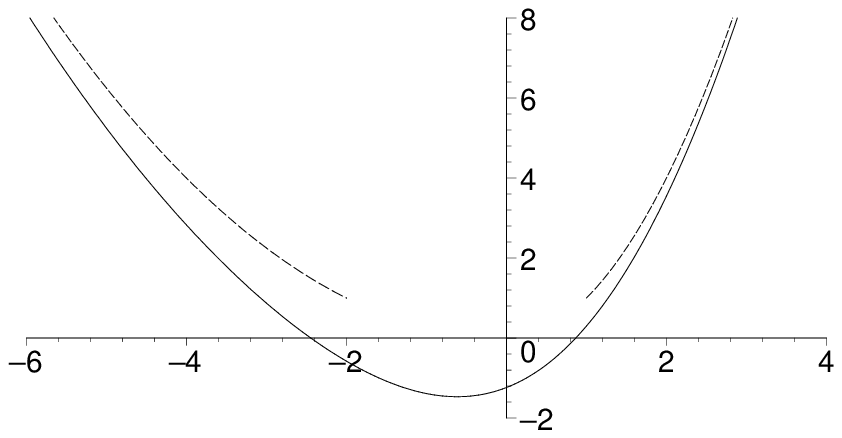}}
\put(0.08,0.3){\includegraphics[width=0.25\textwidth]{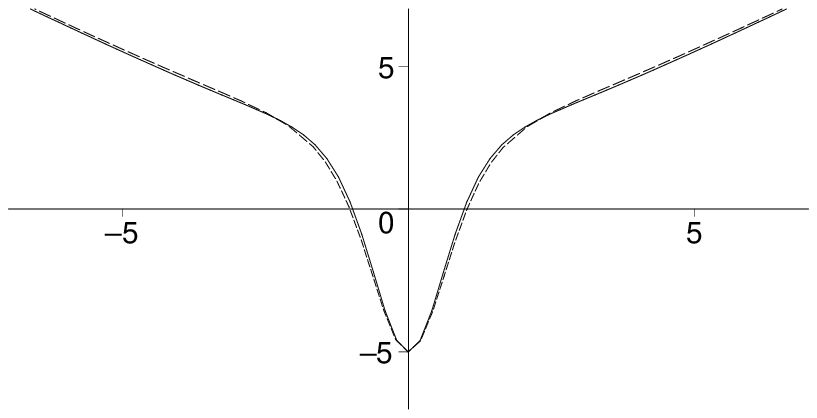}}
\put(0.11,0.25){$s=-5,~ A\sim \frac{25}{4}$}
\put(0.4,0.3){\includegraphics[width=0.25\textwidth]{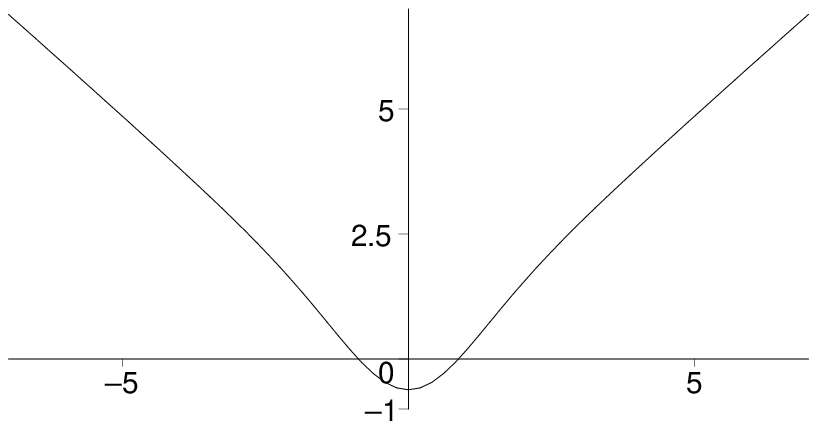}}
\put(0.4,0.25){$s=-0.615, ~A=-1.466$}
\put(0.72,0.3){\includegraphics[width=0.25\textwidth]{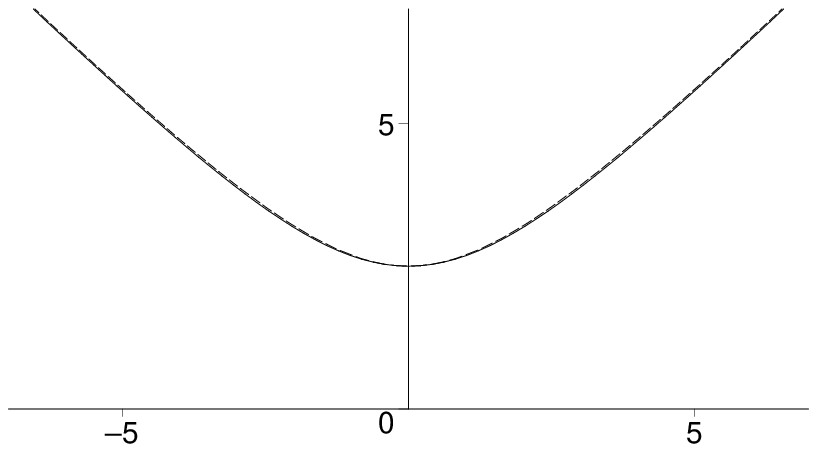}}
\put(0.78,0.25){$s=2.5, ~A\sim \frac{25}{4}$}
\put(0.57,0.5){ {\Large $A$}}
\put(0.93,0.15){ {\Large $s$}}
\end{picture}
}
\end{center}
\caption{Plot of $A$ versus $s=U(0)$ showing the fold point for the
core problem (\ref{core}). The inserts show the solution $U$ versus
$y$ at the parameter values as indicated. The middle insert shows $U$
at the fold point. The dotted lines are the limiting approximations of
$A$ versus $U^{\pm}(0)$ in (\ref{U+-}). See Theorem \ref{thm:core} for more
details.}%
\label{fig:sA}%
\end{figure}}

{ \textbf{Proof.} The proof consists of four steps. In Step 1 we use
formal asymptotics to show that when $A\gg1$, there are exactly two
possible solutions with $U^{\prime}>0$ for $y>0$, as given by
(\ref{U+-}). In Step 2 we rigorously show that there are no solutions
when $-A$ is large enough. In Step 3 we show that the solution branch
with $U^{\prime }>0,\ y>0$ cannot connect with any branch for which
$U^{\prime}\leq0$ at some $y>0.$ It then follows that the two branches
$U^{+}$ and $U^{-}$ must connect to each other. In Step 4 we show that
$\Phi$ is positive and that the fold point $s_{c}$ is unique. }

{\bf Step 1:}. We first consider the case $A\gg1.$ After rescaling
$U=\sqrt{A}v,$ \ $y=\alpha t$ for some $\alpha$ to be determined, 
(\ref{core}) becomes
\[
\frac{1}{\alpha^{2}\sqrt{A}}v_{tt}=v^{2}-1-\frac{\alpha^{2}}{A}y^{2}.
\]
If we choose $\alpha=\sqrt{A}$ then the leading-order equation for $v$
becomes $v^{2}-1-t^{2}\sim0$. This yields $v\sim\sqrt{1+t^{2}}.$ The other
possible choice is $\alpha=A^{-1/4}$, which yields the leading-order 
equation 
\[
v_{tt}=v^{2}-1 \,,
\]
with $v^{\prime}\left(  0\right)  =0,$ $v\left(  t\right)  \sim1$ for
large $t.$ This ODE admits exactly two monotone solutions satisfying
$v\left(  t\right)  \rightarrow1$ as $t\rightarrow\infty.$ These solutions
are given by
\[
v=1 \qquad \mbox{and} \qquad v=1-3\ \operatorname{sech}{}^{2}\left(  \frac
{t}{\sqrt{2}}\right) \,,
\]
which correspond to the inner expansion of $U^{+}$ and $U^{-}$, respectively.
Matching the inner and outer expansion into a uniform solution yields
(\ref{U+})\ and (\ref{U-}).

{\bf Step 2:} Next we show the non-existence of a solution to the
core problem when $-A$ is positive and sufficiently large. To show this,
we rescale
\[
u=\sqrt{-A}v,\ \ \ \ y=\left(  -A\right)  ^{-1/4}t \,.
\]
From (\ref{core}), we obtain
\begin{equation}
v^{\prime\prime}=v^{2}+1-\eps t^{2} \,, \qquad
\eps\equiv\left(  -A\right)^{-3/2} \,. \label{25nov5:19}%
\end{equation}
We will show that no solution to (\ref{25nov5:19}) exists when $\eps>0$
is small enough. First we choose any $a\in(0,1)$ and define $T$ by
\[
T\equiv\sqrt{\frac{1-a}{\eps}} \,.
\]
Then, for $0<t<T$ we have
\begin{equation*}
v^{\prime\prime}   >v^{2}+a \,; \qquad v^{\prime}\left(  0\right) =0 \,, 
 \quad v\left(  0\right)  =v_{0} \,.
\end{equation*}
In particular, $v^{\prime}>0$ for all $t\in(0,T).$ First, we suppose that
$v\left(  0\right) =v_0 \leq0$. Then, under this assumption, we derive
\begin{equation*}
\frac{v^{\prime2}}{2}   \geq\frac{1}{3}v^{3}+av-\left(  \frac{1}{3}v_{0}
^{3}+av_{0}\right)  \geq\frac{1}{3}v^{3}+av \,. 
\end{equation*}
The first step is to show that when $\eps$ is sufficiently large,
$v\left( t\right) $ crosses zero at some value $t=t_{1}.$ There two
subcases to consider. For the first subcase, suppose that $v_{0}<-1.$
Then
\[
\frac{v^{\prime2}}{2}\geq\frac{1}{3}v^{3}-\frac{1}{3}v_{0}^{3} \,,
\]
so that
\[
t_{1}\leq\int_{v_{0}}^{0}\frac{dv}{\sqrt{\frac{1}{3}v^{3}-\frac{1}{3}v_{0}%
^{3}}}\leq C\left\vert v_{0}\right\vert ^{-1/2}\leq C \,.
\]
Therefore, by choosing $\eps$ small enough so that $T\geq C,$ we have
$t_{1}\in\lbrack0,T].$ For the subcase $v_{0}>-1,$ we have $v^{\prime\prime
}\geq a,$ \ $v\geq\frac{a}{2}t^{2}+v_{0}$ so that $t_{1}\leq\sqrt{\frac{2}{a}%
}$. \ Therefore, $t_{1}\in\lbrack0,T]$ by choosing $\eps$ small enough
so that $T\geq\sqrt{\frac{2}{a}}.$ 

{ The second step is to show that $v$ blows up for some $T_{b}\in\left(
0,T\right)  $, provided that $T$ is large enough. This would yield a
contradiction. Indeed we have $\frac{v^{\prime2}}{2}\geq\frac{1}{3}v^{3}+av$
for $v\geq 0$ so that
\[
T_{b}\leq I_{2}+t_{1},\ \ \ \ I_{2}=\int_{0}^{\infty}\frac{dv}{\sqrt{2}
 \sqrt{\frac{1}{3}v^{3}+av}}<\infty.
\]
Therefore a contradiction is attained by choosing $\eps$ so small that
$T>I_{2}+t_{1}$. }

 Finally, if $v_{0}>0$ let $\eta=v_{0}+Bt^{2}.$Then, for large enough $B$ we
have $\eta^{\prime\prime}-\eta^{2}-1+\eps t^{2}\leq0$, so that by a
comparison principle, $v\geq v_{0}+Bt^{2}$ for all $t>0.$ But this is
impossible since we must have $v\rightarrow\sqrt{\eps}t$ for large
values of $t.$ This shows that no solution to the core problem
(\ref{core}) can exist if $-A$ is sufficiently large. 

{\bf Step 3:} We now show that the solution branch with $U^{\prime}>0$
for $y>0$ can never connect to a non-monotone solution branch. We
argue by contradiction. Suppose not. Then consider the first parameter
value $A$ for which a connection occurs. For such a value of $A$,
there must be a point $y_{0}\in [0,\infty)$ such that
$U^{\prime}\left( y_{0}\right) =0$ with $U^{\prime}\geq0$ for any
other $y.$ Suppose first that $y_{0}>0.$ Then we have
$U^{\prime\prime\prime}=2U U^{\prime}-2y$ so that
$U^{\prime\prime\prime}\left( y_{0}\right) =-2y_{0}<0.$ But this
contradicts the assumption that $U^{\prime}\left( y_{0}\right)
=0$ is a minimum of $U^{\prime}.$ If on the other hand $y_{0}=0,$ then
we consider three cases.  First if $U^{\prime\prime}\left( 0\right)
=0$, then from a Taylor expansion we obtain $U\left( 0\right)
=\sqrt{A};$ $U^{\prime}\left( 0\right) =U^{\prime\prime}\left(
0\right) =U^{\prime\prime\prime}\left( 0\right) =0,$ \ $U^{(4)}\left(
0\right) =-2.$ This expansion shows that $U$ is decreasing to the
right of the origin, which contradicts the assumption that $U^{\prime}
\geq0$ for all $y\neq y_{0}.$ Similarly, if $U^{\prime\prime}(0)<0$
then again $U$ is decreasing to the right of the origin, which yields
a similar contradiction. Finally, $U^{\prime\prime}(0)$ cannot be
positive when $y_{0}=0,$ since we assumed that $A$ is the connection
point.

{\bf Step 4:} Define $\Phi(y)$ by
\[
\Phi=\frac{\partial U}{\partial s} \,, \qquad U(0)=s \,.
\]
At the fold point $s=s_c$ where $A^{\prime}\left( s\right) =0$, we
obtain upon differentiating (\ref{core}) that
\begin{equation}
\Phi^{\prime\prime}=2U\Phi \,, \qquad \Phi(0)=1 \,. \label{eigcore}
\end{equation}
To show that $\Phi$ is positive at the fold point, we define $\chi(y)$ by
\[
\chi=\frac{\Phi}{U^{\prime}} \,.
\]
We readily derive that
\begin{equation}
\chi^{\prime\prime}U^{\prime}-2y\chi+2U^{\prime\prime}\chi^{\prime}=0\,.
\label{25nov5:00}%
\end{equation}
Since $\Phi\left( 0\right) =1,$ and $U^{\prime}>0$ for $y>0$, we
obtain that $\chi$ is positive near the origin. In addition, for large
$y$, (\ref{25nov5:00}) reduces to $\chi^{\prime\prime}\sim2y\chi$, which
implies $\chi\rightarrow0$ as $y\rightarrow \infty.$ It follows by the maximum
principle that $\chi>0.$ This shows the positivity of $\Phi.$ Finally,
we establish that the fold point $A^{\prime }\left( s\right) =0$ is
unique. Assuming that $A^{\prime}(s)=0.$, we differentiate
(\ref{core})\ twice with respect to $s$ to obtain
\[
\Phi_{s}^{\prime\prime}=2U\Phi_{s}+2\Phi^{2}-A^{\prime\prime}\left(  s\right)
.
\]
By multiplying both sides of this expression by $\Phi$, and integrating the
resulting expression by parts, we obtain
\[
A^{\prime\prime}\left(  s\right)  =2\frac{\int_{0}^{\infty}
 \Phi^{3} \, dy }{\int_{0}^{\infty} \Phi \, dy} \,.
\]
However, since $\Phi$ is positive then
$A^{\prime\prime}\left(  s\right)  >0$ whenever $A^{\prime}\left(  s\right) 
=0.$ This implies that the fold point is unique. $\blacksquare$ 

\subsection{{The Dimple Eigenfunction}}

Next, we study the qualitative properties of the eigenfunction pair
associated with linearizing (\ref{wsys}) around the steady-state solution
at the fold point where $D=D_c$. We label the steady-state solution at
the fold point $D=D_{c}$ by $u_{c}(x)$ and $w_{c}(x)$. From (\ref{trans2})
and Theorem \ref{thm:core}, we obtain at $D=D_c$ that
\begin{equation*}
    u_{c0}\equiv u_{c}(0)\sim 1 + \eps^{2/3} \mathcal{B}^{1/3} U(0) \,,
  \qquad U(0)=s_c=-0.61512\,. 
\end{equation*}

We linearize (\ref{wsys}) around $u_c$ and $w_c$ by setting
\begin{equation*}
u\left(  x,t\right)    =u_c\left(  x\right)  + e^{\lambda t} \phi(x) \,, \qquad
w\left(  x,t\right)   =w_c\left(  x\right)  + e^{\lambda t} \psi(x) \,.
\end{equation*}
This leads to the eigenvalue problem
\begin{equation}
\lambda\phi   =\eps^{2}\phi_{xx}-f_{u}\left(  u_c,w_c\right)  \phi
-f_{w}\left(  u_c,w_c\right)  \psi \,, \qquad
\frac{\lambda}{\alpha}\left[  \phi+\tau\left(  \psi-\phi\right)  \right]   
=D\psi_{xx}-\beta_{0}\phi \,, \label{weig}
\end{equation}
with $\psi_{x}=\phi_x=0$ at $x=\pm L$ and $D=D_c$.

{ \begin{figure}[tb]
{ \begin{minipage}[t]{0.48\textwidth}
\begin{center}
{\small \setlength{\unitlength}{1\textwidth} \begin{picture}(1,0.5)(0,0)
\put(0.08,0.0){\includegraphics[width=0.9\textwidth]{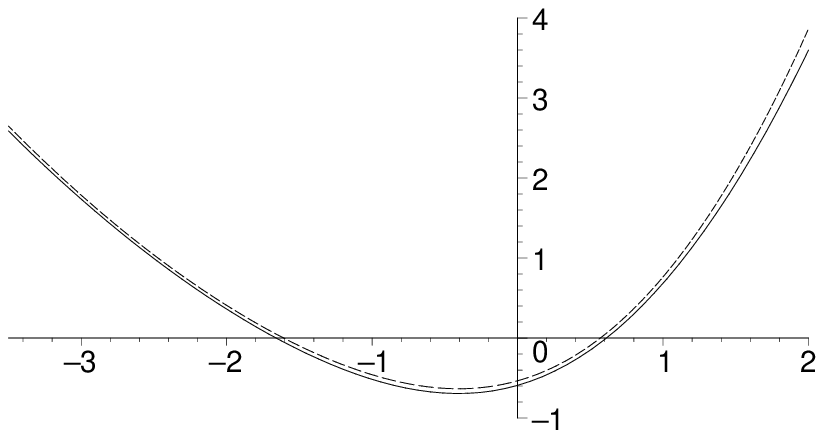}}
\put(0.52,0.4){ $w_1$}
\put(0.93,0.12){ $u_1$}
\end{picture}
(a)
}
\end{center}
\end{minipage}
\begin{minipage}[t]{0.48\textwidth}
\begin{center}
{\small \setlength{\unitlength}{1\textwidth} \begin{picture}(1,0.5)(0,0)
\put(0.08,0.0){\includegraphics[width=0.9\textwidth]{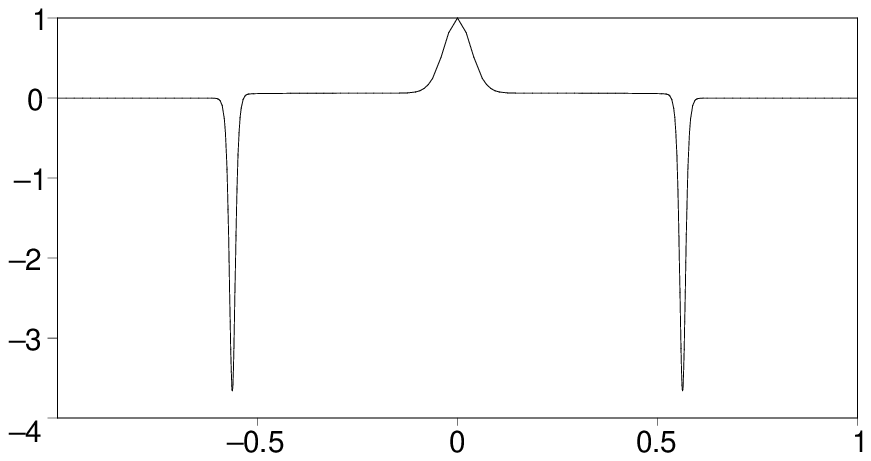}}
\end{picture}
(b)
}
\end{center}
\end{minipage}
}\caption{(a) The plot of $u_{1}(0)$ versus $w_{1}(0)$ computed from
(\ref{trans}). The solid and dashed lines correspond to
numerical computations of the full steady-state system (\ref{ssuw}) and
the core problem (\ref{core}), respectively. (b) The dimple eigenfunction
$\phi$ at the fold point with parameter values $\eps=0.005$,
$\beta_{0}=1.5$ and $D$ is determined numerically as a function of
$u_{1}(0)$}%
\label{fig:u1w1}%
\end{figure}}

Let $u_{0}=u(0)$ and $D=D(u_0)$. Then, if we define $\phi$ and $\psi$ by
\begin{equation}
\phi=\left.  \frac{\partial}{\partial u_{0}}u\left(  x\right)  \right\vert
_{u_{0}=u_{c0}}\,, \qquad \psi=\left.  \frac{\partial}{\partial u_{0}}w\left(
x\right)  \right\vert _{u_{0}=u_{c0}} \,, \label{30nov8:21}%
\end{equation}
it follows from (\ref{weig}) and $D^{\prime}\left(  u_{c0}\right)  =0$ that
(\ref{30nov8:21}) is an eigenfunction pair corresponding to $\lambda=0$. 

We now construct an asymptotic approximation to this eigenpair $\phi$,
$\psi$ of (\ref{weig}) corresponding to $\lambda=0$. In particular, we
show that $\phi$ is an even function that has a dimple shape when
$D=D_c$. This is shown below to be a consequence of the positivity of
the function $\Phi$ in (\ref{eigcore}), together with the integral
constraint $\int_{0}^{L}\phi\, dx=0$, which is readily obtained from
(\ref{weig}). We normalize this eigenfunction by imposing that
$\phi(0)=1$. For $\eps\ll 1$ and $D=D_c$, our analysis below shows
that the asymptotic structure of $\phi$ has four distinct regions: an
inner region of width $O(\eps^{2/3})$ near $x=0$ where $\phi=O(1)$; an
outer region on $x\in (0,l)$ where $\phi=O(\eps^{2/3})$; an inner
region of width $O(\eps)$ near $x=l$ where $\phi=O(\eps^{-1/3})$; and
an outer region on $x\in (l,L]$ where $\phi=0(\eps)$. The first three
regions give asymptotically comparable contributions of order
$O(\eps^{2/3})$ to the integral constraint $\int_{0}^{L}\phi \, dx=0$,
whereas the contribution from the fourth region can be neglected. For
a particular set of parameter values the resulting dimple-shape of the
eigenfunction $\phi$ at $D=D_c$ is shown in Fig.~\ref{fig:u1w1}(b).

We now give the details of the asymptotic construction of $\phi$. We begin
with the internal layer region of width $O(\eps^{2/3})$ near $x=0$. In
this region, we use (\ref{fdef}) and (\ref{trans1}) to calculate
\begin{equation}
  f_u=1+3u_c^2 -2w_c u_c \sim 2\eps^{2/3} \mathcal{B}^{1/3} U_{c} \,; \qquad
  f_w=-u_c^2\sim -1 \,.  \label{reg1}
\end{equation}
Here $U_{c}(y)$ is the solution to the core problem (\ref{core}) at 
the fold point location $A^{\prime}(s_c)=0$ where $D\sim D_c$. Using
(\ref{reg1}) in (\ref{weig}) with $\lambda=0$, we obtain the 
following leading-order system on $0\leq y<\infty$:
\begin{equation}
  \eps^2 \phi_{xx} - 2\eps^{2/3} \mathcal{B}^{1/3} U_{c} \phi + \psi =0 \,;
  \qquad D_c \psi_{xx}-\beta_0 \phi=0 \,, \label{reg1_1}
\end{equation}
with normalization condition $\phi(0)=1$ and with $\phi_x=\psi_x=0$ at
$x=0$. We then introduce the inner variables
$y=x/(\eps^{2/3}\mathcal{B}^{-1/6})$, $\phi=\Phi(y)$, and
$\psi=\Psi(y)$.  Then, (\ref{reg1_1}) becomes
\begin{equation}
  \Phi^{\prime\prime} - 2U_c \Phi + \eps^{-2/3} \mathcal{B}^{-1/3} \Psi=0 \,; 
 \qquad D_c\Psi^{\prime\prime}=\beta_0\eps^{4/3}\mathcal{B}^{-1/3} \Phi \,.
 \label{reg1_2}
\end{equation}
This shows that $\Psi=O(\eps^{4/3})$, and hence $\eps^{-2/3}\Psi=
O(\eps^{2/3})\ll 1$. Therefore, to leading-order, we obtain that
$\Phi$ satisfies (\ref{eigcore}) of Theorem \ref{thm:core} at $s=s_c$,
and that
\begin{equation}
   \Psi\sim\frac{\beta_{0}}{D_c}\eps^{4/3}\mathcal{B}^{-1/3}\Psi_0 \,; \qquad
   \Psi_{0}^{\prime\prime}=\Phi \,, \quad \Psi_{0}^{\prime}(0)=0 \,.
\end{equation}
Therefore, as $y\to \infty$, we have $\Psi_{0}^{\prime}\sim
\left(\int_{0}^{\infty} \Phi(y) \, dy\right)$. Writing the far-field
expansion of $\Psi$ in terms of the outer $x$-variable, we obtain the
far-field matching condition
\begin{equation}
   \Psi \sim \frac{\beta_{0}}{D_c}\eps^{2/3}\mathcal{B}^{-1/6}
 \left(  \int_{0}^{\infty} \Phi(y) \, dy  \right) x \,. \label{reg1_3}
\end{equation}

Next, we analyze the inner region of width $O(\eps)$ near the transition
layer at $x=l$. We introduce the inner variables $y_l={(x-l)/\eps}$,
$\phi=\Phi_l(y)$ and $\psi=\Psi_{l}(y)$. From (\ref{weig}) with $\lambda=0$,
we obtain on $-\infty<y_l<\infty$ that
\begin{equation}
\Phi_{l}^{\prime\prime} -f_{u}\left(U_0,W_0\right) \Phi_l -
 f_{w}\left(U_0,W_0\right)\Psi_l =0 \,, \qquad \Psi_{l}^{\prime\prime}=0
 \,. \label{reg2}
\end{equation}
Here $U_0$ and $W_0$, given in (\ref{U0W0}), satisfy the leading-order 
steady-state inner problem (\ref{U0}). Upon comparing (\ref{U0}) and
(\ref{reg2}), we conclude that $\Phi_l$ is proportional to $U_{0}^{\prime}$
and that $\Psi_l=0$. By using (\ref{U0W0}) for $U_{0}^{\prime}$, we get
\begin{equation}
 \phi \sim \Phi_{l} = c \,\, \mbox{sech}^{2}\left( \frac{x-l}{2\eps}\right)\,,
  \qquad \psi\sim \Psi_l=0 \,. \label{reg2_1}
\end{equation}
Here $c$ is an unknown constant, possibly depending on $\eps$, that is
found below by the global constraint $\int_{0}^{L}\phi\,dx=0$.

In the outer region $x\in (l,L]$, where $u_c=0$, we obtain the
leading-order solution $\phi=\psi=0$. A higher-order construction,
which we omit, shows that $\psi=\phi=O(\eps)$ in this near-boundary
region.  In contrast, in the outer mesa plateau region $x\in (0,l)$,
we set $\lambda=0$ in (\ref{weig}) to obtain $\phi =
-\left[{f_{w}(u_c,w_c)/f_{u}(u_c,w_c)}\right]\psi$ and
$D\psi_{xx}-\beta_{0}\phi = 0$. Then, by using the solution $u_c$ and
$w_c$ to (\ref{outer}) at $D=D_{c}$, we obtain
${f_{w}/f_{u}}=-{u_c^2/(u_c^2-1)}$.  The boundary conditions for
$\psi$ as $x\to 0$ and at $x=l$ are obtained by matching to
(\ref{reg1_3}) and (\ref{reg2_1}), respectively. In this way, we
obtain the following formulation of the leading-order outer problem for 
$\psi$ and $\phi$ on $x\in (0,l)$: 
\bsub \label{reg3m}
\begin{equation}
 D_c\psi_{0xx} - \frac{\beta_0 u_c^2}{u_c^2-1} \psi_0 =0 \,, \quad
  x\in (0,l) \,; \qquad D_c \psi_{0x} \to \beta_0  \quad \mbox{as} 
  \quad x\to 0^{+} \,, \quad \psi_{0}(l)=0 \,. \label{reg3}
\end{equation}
In terms of $\psi_0$, we have
\begin{equation}
   \psi \sim \eps^{2/3} \mathcal{B}^{-1/6} \left(\int_{0}^{\infty} \Phi \, dy
   \right)\psi_0 \,, \qquad  \phi \sim \eps^{2/3} \frac{u_c^2}{u_c^2-1} 
  \mathcal{B}^{-1/6} \left(\int_{0}^{\infty} \Phi \, dy  \right) \psi_0 \,. 
  \label{reg3_c}
\end{equation}
\esub 
By the maximum principle the solution $\psi_0$ to (\ref{reg3}),
which depends only on $\beta_0$, satisfies $\psi_0>0$ on $x\in
(0,l)$. Therefore, since $\Phi>0$ from Theorem \ref{thm:core}, we
conclude that $\phi=O(\eps^{2/3})$ with $\phi>0$ on $x\in (0,l)$ and
$\phi(l)=0$.

Finally, we use the global condition $\int_{0}^{L}\phi\, dx=0$ to 
calculate the constant $c$ in (\ref{reg2_1}). Upon using $\phi\sim \Phi$
for $x=O(\eps^{2/3})$, together with (\ref{reg2_1}) and (\ref{reg3_c}) for
$\phi$ in the plateau and transition regions, we estimate
\begin{align*}
  \int_{0}^{L}\phi \, dx & = \eps^{2/3} \mathcal{B}^{-1/3} 
 \left(\int_{0}^{\infty} \Phi \, dy \right) + \eps^{2/3} 
 \mathcal{B}^{-1/6} \left( \int_{0}^{\infty} \Phi \, dy \right) \left(
  \int_{0}^{l}  \frac{u_c^2}{u_c^2-1} \psi_0 \, dx \right) + \eps c
  \int_{-\infty}^{\infty} \mbox{sech}^2\left({y/2}\right) \, dy \,, \\
        0 &= \eps^{2/3} \mathcal{B}^{-1/6} \int_{0}^{\infty} \Phi \, dy
      \left( 1 + \int_{0}^{l}  \frac{u_c^2}{u_c^2-1} \psi_0 \, dx \right)
      + 4 \eps c \,.
\end{align*}
Therefore, the constant $c$ in (\ref{reg2_1}) satisfies
\begin{equation}
 c \sim c_{0} \eps^{-1/3} \,, \qquad c_0 \equiv -\frac{1}{4} \mathcal{B}^{-1/6}
   \left(\int_{0}^{\infty} \Phi \, dy \right)\left[1 + I(\beta_0) \right] \,,
  \qquad I(\beta_0)\equiv \int_{0}^{l}  \frac{u_c^2}{u_c^2-1} \psi_0 \, dx 
  \,. \label{c0def}
\end{equation}
Here $\mathcal{B}>0$ is defined in (\ref{calB}).  Since
$\int_{0}^{\infty}\Phi \, dy>0$ by Theorem \ref{thm:core}, and
$\psi_0>0$ on $x\in (0,l)$, we get that $c_0<0$. Numerically, we
compute from (\ref{eigcore}) that $\int_{0}^{\infty}\Phi \, dy\approx
1.1857$.  Alternatively, $I(\beta_0)$ must be calculated
numerically from the solution to (\ref{reg3}). We remark that the
integrand in $I(\beta_0)$ is well-defined as $x\to 0$, since although
$u_c\to 1$ as $x\to 0^{+}$, we have $\psi_0\sim {\beta_0 x/D}$ as
$x\to 0^{+}$ to cancel the apparent singularity in the integrand.

We summarize the asymptotic construction of the dimple eigenfunction as
follows:

\begin{proposition}
[Nishiura-Ueyama's Condition 2: Dimple eigenfunction]{ \label{thm:cond2} 
Consider a single-mesa steady-state solution at the fold point
$D=D_c$. Let $\phi$ be the corresponding eigenfunction. For $x=O(\eps^{2/3})$,
we have
\begin{equation*}
\phi\sim\Phi\left(  \mathcal{B}^{1/6}\eps^{-2/3}x\right)\,,\qquad \Phi(0)=1\,.
\end{equation*}
Here $\mathcal{B}={(\beta_{0}-1)/(2D_c)}>0$ and
$\Phi\left(y\right)$, defined in (\ref{eigcore}) of Theorem
\ref{thm:core} at $s=s_c$, is a strictly positive function that decays
at infinity. Alternatively, in an $O(\eps)$ region near $x=l$, we have
\begin{equation*}
\phi\sim c_{0} \,\eps^{-1/3} \operatorname{sech}{}^{2}\left(\frac{x-l}{2\eps}
\right) \,,
\end{equation*}
where $c_0$ is the negative constant, independent of $\eps$, given in
(\ref{c0def}). In the outer plateau region $0<x<l$, then
$\phi=O(\eps^{2/3})$ is determined from (\ref{reg3m}), and this outer
approximation for $\phi$ has a unique zero crossing at $x=l$. This
establishes the dimple-shape of $\phi$ when $\eps\ll 1$.}
\end{proposition}

\subsection{{Universality of the Core Problem\label{sec:universal}}}

In this section we show that the core problem can be derived for a
class of reaction-diffusion systems that have steady-state mesa
solutions. On $x\in [-L.L]$, we begin by constructing a single mesa
steady-state solution for
\begin{equation}
  u_t = \eps^2 u_{xx} + a(u,v) \,, \qquad \sigma v_t = D v_{xx} - v + b(u,v)\,;
\qquad u_x(\pm L,t)=v_{x}(\pm L,t) =0 \,. \label{un:rd}
\end{equation}
We assume that there exists three roots to $a(u,v)=0$ on the interval
$0<v<v_m$ at $u=0$, $u=u_{-}(v)$, and $u=u_{+}(v)$, with
$0<u_{-}(v)<u_{+}(v)$. Furthermore, we assume that
\bsub \label{un:ass}
\begin{equation}
   a_{u}(0,v)<0 \,, \qquad a_{u}(u_-,v)>0 \,, \qquad a_{u}(u_+,v)<0 \,,
\quad \mbox{for} \,\,\, 0<v<v_m \,. \label{un:ass1}
\end{equation}
We write the two roots $u=u_{\pm}(v)$ on $0<v<v_m$ as $v=h(u)$. When
$v=v_m$ the two roots are assumed to coalesce so that $u_m\equiv
u_{-}(v_m)=u_{+}(v_m)$ and $v_m=h(u_m)$. Furthermore, we assume that
there exists a unique value $v_c$ with $0<v_c<v_m$ such that the
Maxwell line condition
\begin{equation}
    \int_{0}^{u_c} a(u,v_c) \, du = 0 \,, \qquad u_c\equiv u_{+}(v_c) 
   \label{un:maxwell}
\end{equation}
is satisfied.  We also assume that $h^{\prime}(u)<0$ for $u>u_m$ and
$h^{\prime}(u)>0$ for $u<u_m$. With these assumptions on $a(u,v)$, we
conclude at the coalescence point that
\begin{equation}
  a_{uu}^{0}\equiv a_{uu}(u_m,v_m)<0 \,, \qquad a_{v}^{0} \equiv 
  a_{v}(u_m,v_m)<0 \,. \label{un:ass2}
\end{equation}
For the function $b(u,v)$ in (\ref{un:rd}) we will assume that
\begin{equation}
    b(0,v)=0 \,; \qquad g(u)\equiv h(u)- b\left[u,h(u)\right] <0 \, \quad
  \mbox{for} \,\,\, u>u_m \,. \label{un:ass3}
\end{equation}
\esub

\begin{figure}[htbp]
\begin{center}
\subfigure[$u_{\pm}(v)$]
{\includegraphics[width = 8cm,height=5.0cm,clip]{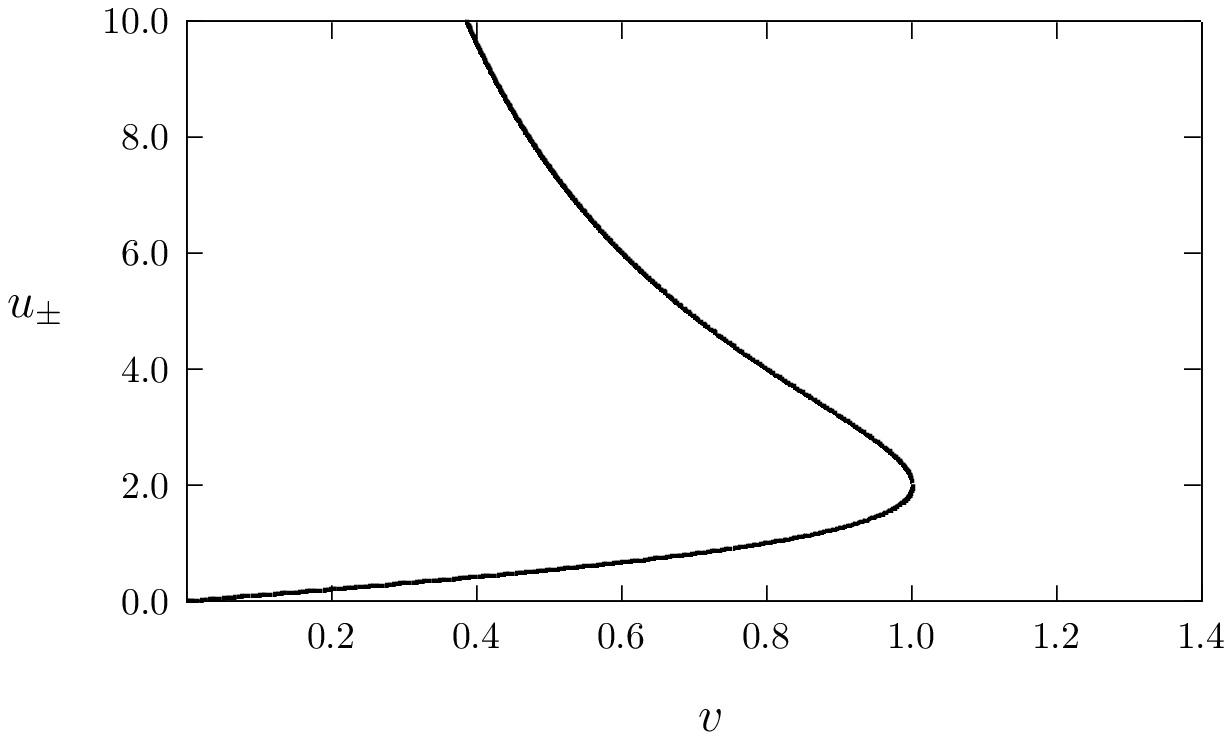} 
\label{un:figu}}
\subfigure[$v=h(u)$]
{\includegraphics[width = 8cm,height=5.0cm, clip]{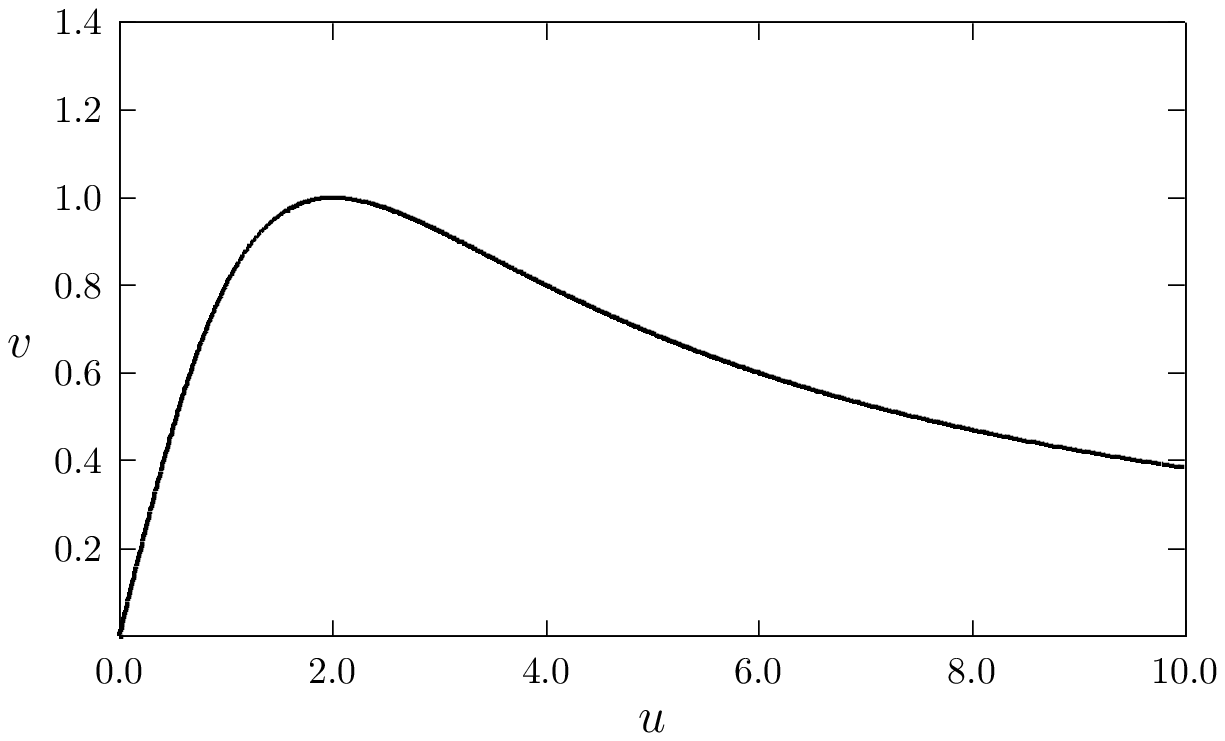} 
\label{un:figv}}
\caption{Left figure: Plot of $u_{\pm}(v)$ from (\ref{usat:upm}) for
the Gierer-Meinhardt model with saturation parameter $k=0.25$.
Right figure: corresponding plot of the inverse function $v=h(u)$ from
(\ref{usat:upm})}
\label{un:figuv}
\end{center}
\end{figure}

A specific example of (\ref{un:rd}) is the Gierer-Meinhardt model with
saturation where $a(u,v)=-u+{u^2/[v(1+ku^2)]}$ with $k>0$, and
$b(u,v)=u^2$.  For this system we calculate
\begin{equation}
   u_{\pm}(v) = \frac{1}{2kv}\left[1 \pm \sqrt{1-4k v^2}\right] \,, \qquad
   v=h(u)=\frac{u}{1+ku^2} \,, \qquad h^{\prime}(u)=\frac{1-ku^2}{(1+ku^2)^2}
 \,. \label{usat:upm}
\end{equation}
Hence, $v_m={1/[2\sqrt{k}]}$, $u_m={1/\sqrt{k}}$, and
$h^{\prime}(u)<0$ for $u>u_m$.  In Fig.~\ref{un:figu} we plot
$u=u_{\pm}(v)$ and in Fig.~\ref{un:figv} we plot $v=h(u)$. The
Maxwell-line condition (\ref{un:maxwell}) is satisfied when
(cf.~\cite{kww-stripe})
\begin{equation}
   v_c = \frac{0.4597}{\sqrt{k}} \,, \qquad u_c\equiv u_{+}(v_c) = 
  \frac{1.515}{\sqrt{k}} \,. \label{usat:max}
\end{equation}
In addition, we calculate from (\ref{un:ass3}) that
\begin{equation}
   g(u) = \frac{u}{1+ku^2} \left[1-u(1+ku^2)\right] \,. \label{usat:g}
\end{equation}
Since $g^{\prime}(u)<0$ for $u>u_m={1/\sqrt{k}}$, and 
$g\left({1/\sqrt{k}}\right)<0$ when $0<k<4$, we have $g(u)<0$ for $u>u_m$
when $0<k<4$.

We now return to the general case under the assumptions (\ref{un:ass})
and we construct a single mesa steady-state solution of the type shown
Fig.~\ref{fig:single-mesa}. We first derive an expression for the
critical value $D_c$ of $D$ for which no single mesa steady-state
solution exists when $D<D_c$.

Near the interface at $x=l$ we introduce the inner expansion
\begin{equation}
 u  =U_{0}\left(  y\right)  + \eps U_{1}\left(  y\right)
+\ldots\,,\ \ \ \ \ v=V_{0}+\eps V_{1}\left(  y\right)  +\ldots \,, \qquad
  y=\eps^{-1}(x-l) \,. \label{un:inn}
\end{equation}
From the steady-state problem for (\ref{un:rd}), we obtain
\bsub \label{un:sys}
\begin{gather}
  U_{0}^{\prime\prime} + a\left(U_0,V_0\right)=0 \,, \qquad  
   V_{0}^{\prime\prime}=0 \,, \label{un:U0} \\
  U_{1}^{\prime\prime}+ a_u(U_{0},V_{0}) U_1 = -a_v(U_0,V_0)V_1 \,, \qquad
  V_{1}^{\prime\prime}=0 \,. \label{un:U1}
\end{gather}
\esub

The solution to the leading-order problem is $V_0=v_c$, where $v_c$
satisfies (\ref{un:maxwell}), and $U_0(y)$ is the unique heteroclinic 
connection satisfying
\begin{equation}
  U_{0}(-\infty)=u_{+}(v_c)=u_c \,, \qquad U_{0}(\infty)=0 \,, \qquad
  U_{0}(0)={u_c/2} \,. \label{un:het}
\end{equation}
At next order we obtain that $V_{1}=V_{11}y+ V_{12}$, for some
constants $V_{11}$ and $V_{12}$. The solvability condition
for (\ref{un:U1}) determines $V_{12}$ in terms of $V_{11}$ as
\begin{equation*}
  V_{12} \int_{-\infty}^{\infty} a_v(U_0,v_c) \, U_{0}^{\prime}\, dy=-V_{11} 
 \int_{-\infty}^{\infty} a_{v}(U_0,v_c) \, y \, U_{0}^{\prime} \, dy \,.
\end{equation*}
Then, by matching to the outer solution for $v$ we obtain
$V_{11}=v^{\prime}(l^{\pm})$.

The outer problems for $v$ determine $v^{\prime}(l^{\pm})$. In the
mesa region $0\leq x \leq l$, where $v=h(u)$, we readily derive the
following outer problem
\begin{equation}
 Dv^{\prime\prime}=g\left(u\right) \,, \qquad 0<x<l \,; \qquad
 v\left(  l\right)  = v_c \,, \qquad v^{\prime}\left(  0\right)=0 \,.
 \label{un:outer_m}
\end{equation}
Here $g(u)$ is defined in (\ref{un:ass3}). The corresponding $u$ is
given by $u(x)=u^{+}[v(x)]$ with $u(l)=u^{+}(v_c)\equiv u_c$. We
require that $0<v<v_m$ at each $x\in (0,l)$ so that $u>u_m$ on $x\in
(0,l)$. In contrast, since $b(0,v)=0$ by (\ref{un:ass3}), we obtain in
the outer region $l\leq x\leq L$ that $u=0$ and that
\begin{equation}
 Dv^{\prime\prime}=v \,, \qquad l<x<L \,; \qquad
 v\left(  l\right)  = v_c \,, \qquad v^{\prime}\left(  L\right)=0 \,.
 \label{un:outer_f}
\end{equation}
Since $V_{11}$ is a constant, the solutions to (\ref{un:outer_m}) and
(\ref{un:outer_f}) are joined by the condition that
$v^{\prime}(l^{-})=v^{\prime}(l^{+})$.

The reduction of (\ref{un:outer_m}) and (\ref{un:outer_f}) to a
quadrature relating $u(0)\equiv u_0$ to the length of the domain $L$
is very similar to that done for the Brusselator. We first multiply
(\ref{un:outer_m}) by $v^{\prime}=h^{\prime}(u) u^{\prime}$ and integrate
to get
\begin{equation*}
 D\frac{v^{\prime2}}{2}=F\left(  u;u_0\right) \,, \qquad
F\left(  u;u_{0}\right)  \equiv\int_{u_{0}}^{u}g\left(  s\right)  h^{\prime
}(s) \, ds \,. 
\end{equation*}
Since $h^{\prime}(u)<0$ and $g(u)<0$ for $u>u_m$ (see
(\ref{un:ass3})), we obtain that $F(u;u_0)>0$ for $u>u_0$. By taking
the negative square root, we calculate
\begin{equation}
 \frac{dv}{dx} = -\sqrt{ \frac{2}{D} } \sqrt{ F(u;u_0) }  <0 \,, \qquad
\frac{du}{dx}=  -\sqrt{ \frac{2}{D} } \frac{ \sqrt{F\left(  u;u_0\right)}}
 { h^{\prime}(u) }   >0 \,. \label{un:uone}
\end{equation}
By integrating (\ref{un:uone}) with $u(0)=u_0$ and $u(l)=u_c=u_{+}(v_c)$, we
obtain a relation between the half-length $l$ of the mesa and $u_0$
\begin{equation}
 -\frac{l}{\sqrt{D}} =\int_{u_{0}}^{u_{+}(v_c)}%
\frac{h^{\prime}\left(  u\right)  }{\sqrt{2F\left(  u;u_0\right)  }}
 \, du = \frac{\sqrt{2F\left( u_{+}(v_c);u_{0}\right)  }}
 {g\left[u_{+}(v_c)\right]} + \int_{u_{0}}^{u_{+}(v_c)} 
 \frac{g^{\prime}\left(  u\right)  }
{ \left[g\left(  u\right)\right]^2}\sqrt{2F\left(  u;u_0\right) } \, du <0 \,.
  \label{un:lval}
\end{equation}
In the outer region $l<x<L$, we solve (\ref{un:outer_f}) to obtain
\begin{equation}
   v(x) = v_c \frac{ \cosh\left[ { (L-x)/\sqrt{D}} \right]}
          {\cosh\left[ { (L-l)/\sqrt{D}} \right]} \,, \qquad
      v_{x}(l^{+})=-\frac{v_c}{\sqrt{D}} \tanh\left[ {(L-l)/\sqrt{D}}
\right] \,. \label{un:flat}
\end{equation}
By setting $v_{x}(l^{-})=v_{x}(l^{+})$, we obtain that
\begin{equation}
   \frac{ (L-l)}{\sqrt{D}} = \tanh^{-1} \left(
   \frac{ \sqrt{2 F(u_{+}(v_c);u_0)}}{v_c} \right) \,, \qquad
  \hbox{when} \quad \frac{ \sqrt{2 F(u_{+}(v_c);u_0)}}{v_c}<1 \,.
 \label{un:Lval}
\end{equation}
Finally, upon combining (\ref{un:lval}) and (\ref{un:Lval}), we obtain
the following expression relating $u_0\equiv u(0)$ to $L$:
\begin{equation}
  \frac{L}{\sqrt{D}} = \chi(u_0) \equiv
\tanh^{-1} \left(\frac{ \sqrt{2 F(u_{+}(v_c);u_0)}}{v_c} \right) -
  \frac{\sqrt{2F\left( u_{+}(v_c);u_{0}\right)  }}
 {g\left[u_{+}(v_c)\right]} - \int_{u_{0}}^{u_{+}(v_c)} 
 \frac{g^{\prime}\left(  u\right)  }
{ \left[g\left(  u\right)\right]^2}\sqrt{2F\left(  u;u_0\right) } \, du \,.
 \label{un:exis}
\end{equation}
Noting that $F_{u_0}(u;u_0)<0$ for $u>u_0$, a simple calculation shows
that $\chi(u_0)$ is a decreasing function of $u_0$ when
$u_0>u_m$. Therefore, for the existence of a single mesa steady-state
solution, we require that $D>D_c$, where $D_c\equiv {L^2/[\chi(u_m)]^2}$.

Next, we show that the core problem (\ref{core}) determines the local
internal layer solution behavior near the origin when $D=D_c$. In the
this layer near $y=0$ we expand
\begin{equation}
  u =u_m + \delta u_{1}+\cdots \,, \qquad v = v_m + \delta^{2} v_{1}+\cdots 
 \,, \qquad z={x/\delta} \,, \qquad D   =D_{c}+\cdots \,, \label{un:layer}
\end{equation}
where $\delta\ll 1$. The nonlinear terms in (\ref{un:rd}) are calculated
as
\begin{equation}
   a(u,v)\sim a^{0} + a_{u}^{0} (u-u_m) + \frac{ a_{uu}^{0}}{2} (u-u_m)^2
    + a_{v}^{0} (v-v_m) + \cdots \sim \delta^{2} \left( \frac{u_{1}^{2}}{2}
  a_{uu}^{0} + a_v^{0} v_1 \right) \,. \label{un:eval}
\end{equation}
Here the superscript $0$ denotes the evaluation of partial derivatives of $a$
at $u=u_m$ and $v=v_m$. In obtaining (\ref{un:eval}) we used 
$a^{0}=a_{u}^{0}=0$. By substituting (\ref{un:layer}) and (\ref{un:eval})
into the steady-state problem for (\ref{un:rd}), and choosing 
$\delta=\eps^{2/3}$, we obtain
\begin{equation}
    u_{1zz} + \frac{a_{uu}^{0}}{2} u_{1}^2 + a_{v}^{0} v_1
 =0 \,, \qquad D_c v_{1zz}    = g(u_m) \,. \label{un:norm1}
\end{equation}
Here $g(u)$, with $g(u_m)<0$, is defined in (\ref{un:ass3}). The solution
for $v_1$ is written as
\begin{equation}
   v_1 = - \mathcal{A} - \mathcal{B} z^2 \,, \qquad
 \mathcal{B} = -\frac{g(u_m)}{2D_c} >0 \,, \qquad 
  \mathcal{A}=-v_{1}\left(  0\right)\,. \label{un:calB}
\end{equation}
Then, (\ref{un:norm1}) for $u_1$ becomes
\begin{equation}
  u_{1zz} + \frac{a_{uu}^{0}}{2} u_{1}^2 - a_{v}^{0} \left(\mathcal{A} +
 \mathcal{B} z^2 \right) = 0 \,. \label{un:norm2}
\end{equation}
From (\ref{un:ass2}) we recall that $a_{uu}^{0}<0$ and $a_{v}^{0}<0$.
Finally, we rescale (\ref{un:norm2}) by  introducing $C$, $\mu$, and $A$,
 by $u_1=C U$, $z=\mu y $, and ${\mathcal A}=\mathcal{B} \mu^2 A$. Then,
(\ref{un:norm2}) is transformed precisely to the core problem (\ref{core}) 
for $U(y)$ and $A$ when
\begin{equation}
\mu = \left( \frac{2}{a_{uu}^{0} a_{v}^0} \right)^{1/6} {\mathcal B}^{-1/6}\,,
  \qquad C = -\frac{2}{a_{uu}^{0}} 
   \left( \frac{2}{a_{uu}^{0} a_{v}^0} \right)^{-1/3} {\mathcal B}^{1/3}\,,
  \qquad \mathcal{A}= {\mathcal B} \mu^2 A \,.\label{un:norm3}
\end{equation}
By combining these transformations, we obtain the following characterization
of the internal layer near the origin:
\bsub \label{un:normlast}
\begin{align}
  u-u_m &\sim \eps^{2/3} C U(y) \sim -\frac{2\eps^{2/3}}{a_{uu}^{0}} 
   \left(  \frac{2}{a_{uu}^0 a_v^0} \right)^{-1/3} \mathcal{B}^{1/3}\, U(y) \,,
   \label{un:normlast_a} \\
  v-v_m &\sim \eps^{4/3} v_1 \sim -\eps^{4/3} \mathcal{B}^{2/3} 
   \left(  \frac{2}{a_{uu}^0 a_v^0} \right)^{1/3} (A + y^2)  \,,
   \label{un:normlast_b} \\
   y &= \frac{x}{\mu\eps^{2/3}}  = \frac{x}{\eps^{2/3}} 
   \left(  \frac{2}{a_{uu}^0 a_v^0} \right)^{-1/6} \mathcal{B}^{1/6} \,.
    \label{un:normlast_c}
\end{align}
\esub
Here $\mathcal{B}$ is defined in (\ref{un:calB}).

Using the result from Theorem \ref{thm:core} for the core problem
(\ref{core}), we conclude that the bifurcation diagram near the
existence threshold of $D$ for a single mesa steady-state solution of
(\ref{un:rd}) has a saddle-node structure. Recall that at the
saddle-node point $U(0)\approx-0.61512<0$ and $A=A_c\approx
-1.46638<0$ (see Theorem \ref{thm:core}). Therefore, from
(\ref{un:normlast}), we have $u(0)<u_m$ and $v(0)-v_m>0$ at the
saddle-node point, as expected.

For the Gierer-Meinhardt model with saturation where
$a(u,v)=-u+{u^2/[v(1+ku^2)]}$, $b(u,v)=u^2$, $u_m={1/\sqrt{k}}$, and
$v_m={1/[2\sqrt{k}]}$, we calculate that
\begin{equation}
   a_v^{0}=-2 \,, \qquad a_{uu}^{0}=-\sqrt{k} \,, \qquad  
   g(u_m) = \frac{1}{2\sqrt{k}} \left[ 1 - \frac{2}{\sqrt{k}}\right] \,
  \quad \mbox{with} \,\,\, 0<k<4 \,.  \label{un:gmsnorm}
\end{equation}
The existence threshold $D_c=D_{c}(k)$ can be computed numerically from
(\ref{un:exis}) for a given domain half-length $L$.

Finally, we remark on the local behavior of the time-dependent solution
to (\ref{un:rd}) in the internal layer region. If we substitute
(\ref{un:layer}) with $u_1=u_1(z,t)$ and $v_1=v_1(z,t)$, we readily
obtain that
\begin{equation}
   \eps^{-2/3} u_{1t} = u_{1zz} + \frac{a_{uu}^{0}}{2} u_1^2 + a_{v}^0 v_1
 \,, \qquad \sigma \eps^{4/3} v_{1t} =D_c v_{1zz} - g(u_m) \,.
\end{equation}
We then introduce $C$ and $\tau$ defined by $t=\eps^{-2/3}\mu^2 \tau$
and $u_1=C U$. In this way we obtain, $\sigma \eps^2 \mu^{-2}v_{1t}=
D_c v_{1zz}-g(u_m)$. Thus, $v_1$ is quasi-steady, and $D_cv_{1zz}=g(u_m)$.
The corresponding equation for $U(y,\tau)$ is
\begin{equation}
   U_{\tau} = U_{yy} - U^2 + A + y^2 \,.  \label{un:blow}
\end{equation}
If we take $A<A_c$ and even initial data $U(y,0)$ with $U(0,0)<s_c$,
which is below the existence threshold for the steady-state core problem,
then (\ref{un:blow}) should exhibit the finite-time blowup $U\to-\infty$ as
$\tau\to T^{-}$. The local structure of the solution near the blowup
point $y=0$ and $\tau=T$ is independent of the lower-order terms $A+y^2$
in (\ref{un:blow}), and is given from \cite{fk} as
\begin{equation}
  U(y,\tau) \sim -(T-\tau)^{-1/2} \left[ 1 + \frac{1}{4|\log(T-\tau)|} - 
  \frac{y^2}{8(T-\tau) |\log(T-\tau)|} \right] \,. \label{un:blow2}
\end{equation}
The analysis leading to (\ref{un:blow}) is, of course, not a valid
description of the solution to (\ref{un:rd}) when $\tau\to T^{-}$
since full nonlinear effects in (\ref{un:rd}) must be accounted for
near the singularity time. However, this analysis does suggest the
formation of a {\emph large amplitude} finger, such as shown in
Fig.~\ref{fig:split}(b), when $D$ is reduced significantly below $D_c$.

\setcounter{equation}{0}
\setcounter{section}{2}
\section{{The Stability of the Mesa Pattern\label{sec:stability}}}

In this section we show that the steady-state $K$-mesa pattern is stable 
when $D>D_{c}$ and $\tau=0$. We linearize (\ref{wsys}) around this steady-state
solution by letting
\begin{equation*}
u\left(  x,t\right)     =u\left(  x\right)  + e^{\lambda t}
\phi(x) \,, \qquad
w\left(  x,t\right)     =w\left(  x\right)  + e^{\lambda t} \psi(x) \,,
\end{equation*}
to obtain (\ref{weig}). Upon setting $\tau=0$ in (\ref{weig}), we
obtain the eigenvalue problem
\begin{equation}
\lambda\phi   =\eps^{2}\phi_{xx}-f_{u}\left(  u,w\right)  \phi
-f_{w}\left(  u,w\right)  \psi \,, \qquad \frac{\lambda}{\alpha}\phi   =
 D\psi_{xx}-\beta_{0}\phi \,, \label{lin}
\end{equation}
where $f(u,w)$ is defined in (\ref{fdef}). The main result of this section 
is as follows:

\begin{theorem}
[Nishiura-Ueyama's Condition 3]{ \label{thm:stability}\bigskip\ Consider a 
symmetric $K$-mesa steady-state solution as constructed in\ Proposition 
\ref{prop:Dc} on a domain of length $2KL,$ with $D>D_{c}.$ When $\tau=0$,
the spectrum of (\ref{lin}) admits only real eigenvalues with
$\lambda<0$.}
\end{theorem}

To show Theorem \ref{thm:stability}, we first reformulate (\ref{lin}) as a
singular limit eigenvalue problem (SLEP) in the limit $\eps\to 0$ with
$\alpha=O\left(  \eps^{2}\right)$, in order to derive a reduced set of
equations, independent of $\eps$, for the eigenvalues. The following Lemma
characterizes this reduced system and its eigenvalues:

\begin{lemma}
[SLEP reduction]{\label{thm:slep-reduction} Let $\lambda$ be an eigenvalue
associated with the $K$-mesa steady-state solution $w,u$ on an interval of 
length $2KL$, with an interface at $x=l,$ as described in Proposition 
\ref{prop:Dc}. The leading $2K$ eigenvalues of the eigenvalue problem 
(\ref{lin}) are of order $O\left(  \alpha\right)  $ with
$\alp=O(\eps^2)$. These eigenvalues are characterized as follows: Define 
$\lambda_1$ by
\[
\lambda=\alpha\lambda_{1} \,, %
\]
where $\alp=O(\eps^2)$, and let $u_{e}$, $u_{o}$ be the solutions of
the differential equation
\[
D\psi^{\prime\prime}-\left(  \beta_{0}+\lambda_{1}\right)  \frac{u^{\prime}%
}{w^{\prime}}\psi=0 \,,
\]
satisfying the boundary conditions
\begin{equation*}
u_{o}\left(  0\right)     =0\,,\quad u_{o}^{\prime}\left(  l\right)  =1 \,;
 \qquad u_{e}^{\prime}\left(  0\right) =0 \,, \quad u_{e}^{\prime}\left(
  l\right) =1 \,.
\end{equation*}
Then $\lambda_{1}$ satisfies
\begin{equation}
-\frac{1}{\sigma}\left(  \beta_{0}+\lambda_{1}\right)  =\frac{L-l}{\sqrt{2}}\,,
 \label{lam1eq}%
\end{equation}
where $\sigma$ is one of the eigenvalues of the $2K \times 2K$ matrix
\begin{equation}
M = \left[
\begin{array}
[c]{ccccccc}%
-\frac{b}{\delta} & \frac{a}{\delta} &  &  &  &  & \\
\frac{a}{\delta} & -\frac{b}{\delta}-\frac{1}{2d} & \frac{1}{2d} &  &  &  & \\
& \frac{1}{2d} & -\frac{b}{\delta}-\frac{1}{2d} & \frac{a}{\delta} &  &  & \\
&  & \frac{a}{\delta} & -\frac{b}{\delta}-\frac{1}{2d} & \ddots &  & \\
&  &  & \ddots & \ddots & \frac{1}{2d} & \\
&  &  &  & \frac{1}{2d} & -\frac{b}{\delta}-\frac{1}{2d} & \frac{a}{\delta}\\
&  &  &  &  & \frac{a}{\delta} & -\frac{b}{\delta}%
\end{array}
\right] \,.  \label{M}%
\end{equation}
The entries of the matrix $M$ are
\begin{equation}
a   =\frac{u_{e}\left(  l\right)  -u_{o}\left(  l\right)  }{2} \,, \qquad
b=\frac{u_{e}\left(  l\right)  +u_{o}\left(  l\right)  }{2} \,, \qquad
\delta   =b^{2}-a^{2} \,, \qquad  d=L-l \,. \label{def:ab}
\end{equation}
The eigenvalues of $M$ are given explicitly by
\begin{equation}
\sigma_{j\pm}    =-\frac{1}{2d}-\frac{b}{\delta}\pm\sqrt{\left(  \frac
{a}{\delta}\right)  ^{2}+\left(  \frac{1}{2d}\right)  ^{2}+\frac{a}{d \delta}
 \cos\left(\frac{\pi j}{K}\right) } \,, \quad  j=1,\ldots, K-1\,; \qquad
\sigma_{0\pm}    =-\left(  \frac{1}{b\pm a}\right) \,.\label{sigma}
\end{equation}
For the special case of one mesa, where $K=1$, there are two small 
eigenvalues corresponding to either an even or an odd eigenfunction. These
eigenvalues satisfy
\begin{equation}
\left(  \beta_{0}+\lambda_{1}\right)  u_{e}\left(  l\right)  =\frac{L-l}%
{\sqrt{2}}\text{ \ \ \ and \ \ \ \ }\left(  \beta_{0}+\lambda_{1}\right)
u_{o}\left(  l\right)  =\frac{L-l}{\sqrt{2}} \,. \label{eig:onemesa}
\end{equation}
}\end{lemma}

This proof of this lemma is given in Appendix A. Here we will use it to
prove Theorem \ref{thm:stability}.

\textbf{Proof of Theorem \ref{thm:stability}.} Define $\mu$ by
\[
\mu=\lambda_{1}+\beta_{0} \,.
\]
From (\ref{w=h(u)}) we obtain on the interval $x\in \left( 0,l\right)$
that $\frac{u^{\prime}}{w^{\prime}}=\frac{u^2}{u^2 -1}>0$ since
$u\in\left( 1,\sqrt{2}\right] $ on this interval. With this
preliminary result, the proof of Theorem \ref{thm:stability} consists
of four steps.

{\bf Step 1:} Let $\mu=\lambda_{1}+\beta_{0}$ and define $f\left(
\mu\right) =\mu u_{o}\left( l\right) .$ In this step we will show that
the function $\mu\rightarrow u_{0}\left( l\right) $ is decreasing
whereas $f\left( \mu\right) $ is increasing for all $\mu>0.$

The former claim is easy to show. Indeed, let $u_{i}$ be a solution of
$u^{\prime\prime}-h_{i}u=0$ with $u_{i}\left( 0\right) =0,\
u_{i}^{\prime }\left( l\right) =1$, for $i=1,2$, and with
$0<h_{1}(x)<h_{2}(x).$ Then, from the comparison principle, we find
that $u_{1}>u_{2}.$ Now take $0<\mu_{1}<\mu_{2}$. Applying this
comparison principle with $h_{1}=\frac{\mu_1}{D}
\frac{u^{\prime}} {w^{\prime}}$
 and $h_{2}=\frac{\mu_2}{D}\frac{u^{\prime}}{w^{\prime}}$, we
immediately find that $u_{o}\left( l;\mu_{1}\right) >u_{o}\left(
l;\mu_{2}\right)$.

Next we show the more difficult result that $f\left( \mu\right) $ is
increasing. Define $h\left( x\right)
=\frac{1}{D}\frac{u^{\prime}}{w^{\prime}}$ so that $u_{0}$ satisfies%
\begin{equation}
u_{o}^{\prime\prime}-\mu h\left(  x\right)  u_{o}    =0\,, \quad h>0\,; \qquad
u_{o}\left(  0\right)     =0\,, \quad u_{o}^{\prime}\left(  l\right)  =1 \,.
 \label{stab0}
\end{equation}
Define $v$ and $v_{\mu}$ by $v=\frac{\partial}{\partial\mu}\left(
\mu u_{0}\right)  $ and $v_{\mu}=\frac{\partial}{\partial\mu}v.$ Then, we 
readily obtain
\begin{align*}
v^{\prime\prime}-\mu hv&=\mu hu_{o}\,; \qquad v\left( 0\right) =0\,,
  \quad v^{\prime}\left( l\right) =1 \,, \\ 
 v_{\mu}^{\prime\prime}-\mu  hv_{\mu} &=2hv\,; 
   \qquad v_{\mu}\left( 0\right) =0\,, \quad v_{\mu}^{\prime}\left( l\right) 
  =0 \,.
\end{align*}
First note that by the maximum principle, $u_{o}>0$ for all
$x\in\left( 0,l\right) $ so that $v^{\prime\prime}-\mu hv>0$ in
$\left( 0,l\right).$ Now suppose that $v(l)\leq0.$ Then, by the
maximum principle, $v<0$ for all $x\in\left( 0,l\right) .$ But this
implies that $v_{\mu}^{\prime\prime}-\mu hv_{\mu}<0$ 
inside $(0,l).$ It then follows
by the maximum principle that $v_{\mu}>0$ for all $x\in(0,l).$ Since
$f^{\prime}\left( \mu\right) =v\left( l\right) $ and
$f^{\prime\prime}\left( \mu\right) =v_{\mu}\left( l\right)$, we conclude
that $f^{\prime\prime}\left( \mu\right) >0$
whenever $f^{\prime}\left( \mu\right) <0.$ It follows that $f$ has no
local maximum.  Therefore, to complete the proof of Step 1, it
suffices to show that $f^{\prime}\left( 0\right) >0.$

For $\mu\ll 1$, the leading-order solution to (\ref{stab0}) is
$u_{o}\left( x\right) \sim x$. It follows that $f\left( \mu\right)
\sim\mu l$ as $\mu\rightarrow0$ so that $f^{\prime}\left( 0\right)
=l>0.$ This completes the proof of Step 1.

{\bf Step 2:} We show that $\sigma_{\min}<\sigma<0$ where%
\begin{equation}
\sigma_{\min}\equiv-\left(  \frac{1}{u_{o}\left(  l\right)  }+\frac{1}
 {d} \right)  \,, \qquad d=L-l \,. \label{sigmin}
\end{equation}
To show this result, we must establish that $0<u_{o}\left( l\right)
 <u_{e}\left( l\right)$, where $u_{o}\left( l\right) =b-a$,
and $u_{e}\left( l\right)=b+a$ from (\ref{def:ab}). This result 
follows from a comparison principle, which yields that
$0<u_{o}\left( x\right) <u_{e}\left( x\right) $ for all 
$x\in\left(0,l\right]$. A simple calculation shows that the result 
$\sigma_{\min}<\sigma<0$ readily follows from (\ref{sigma}) upon using 
$0<u_{o}(l)<u_{e}(l)$. This completes the proof of Step 2.

{\bf Step 3:} Since $\sigma>\sigma_{\min}$, we derive that
\begin{equation}
-\frac{1}{\sigma}\left(  \beta_{0}+\lambda_{1}\right)  > {\cal G}(\mu)
  \equiv \frac{\mu u_{o}\left(
l\right)  }{\frac{1}{L-l}u_{o}\left(  l\right)  +1} \,. \label{10apr4:30}%
\end{equation}
We now calculate ${\cal G}(\beta_0)$ corresponding to $\lambda_1=0$. When
$\lambda_{1}=0$, (\ref{stab0}) becomes
\begin{equation}
u_{o}^{\prime\prime}-\frac{u^{\prime}}{w^{\prime}}\beta_0 u_{o}=0\,; \qquad
u_{o}\left(  0\right)     =0\,, \quad u_{o}^{\prime}\left(  l\right)  =1 \,.
 \label{7:22apr10}
\end{equation}
By differentiating (\ref{eqw}), we note that $w^{\prime}$ satisfies
(\ref{7:22apr10}) on $[0,l]$ with $w^{\prime}\left( 0\right)
=0.$ Therefore, $u_{o}\left( x\right)={w^{\prime}(x)/w^{\prime\prime}(l)}$. 
We then calculate using (\ref{eqw}) that
\[
w^{\prime\prime}\left(  l\right)  =\frac{1}{D}\left(  \beta_{0}\sqrt
{2}-1\right)  \qquad \mbox{and} \qquad 
 w^{\prime}\left(  l\right) =\frac{L-l}{D} 
\,.
\]
Therefore, for $\mu=\beta_{0,}$ we obtain 
$u_{o}\left(  l\right) ={ (L-l)/(\beta_{0}\sqrt{2}-1)}$ and consequently
\[
  {\cal G}(\beta_0)\equiv \frac{\beta_{0}u_{o}\left(  l\right)  }
  {\frac{1}{L-l}u_{o}\left(  l\right) +1}=\frac{L-l}{\sqrt{2}} \quad 
 \mbox{when} \quad \mu=\beta_{0} \,.
\]
Next, by Step 1, we readily find that ${\cal G}(\mu)$ in
(\ref{10apr4:30}) is an increasing function of $\mu$ whenever $\mu > 0$. 
Therefore,
\[
-\frac{1}{\sigma}\left(  \beta_{0}+\lambda_{1}\right)  > {\cal G}(\beta_0)
 \equiv \frac{L-l}{\sqrt{2}} \quad \mbox{for all} \quad \mu>\beta_{0} \,.
\]
We conclude that (\ref{lam1eq}) cannot be satisfied if $\lambda_{1}$ is
real and positive.

{\bf Step 4:} To complete the proof of Theorem \ref{thm:stability}, it
suffices to show that all roots $\lambda_1$ to (\ref{lam1eq}) are purely
real. To do so, we decompose $M$ into the two block-diagonal matrices
\[
M=M_{1}+M_{2} \,,
\]
where%
\[
M_{1}=\left[
\begin{array}
[c]{cccc}%
\frame{$%
\begin{array}
[c]{rr}%
\mathbf{-}\frac{b}{\delta} & \frac{a}{\delta}\\
\frac{a}{\delta} & \mathbf{-}\frac{b}{\delta}%
\end{array}
$} &  &  & \\
& \frame{$%
\begin{array}
[c]{rr}%
\mathbf{-}\frac{b}{\delta} & \frac{a}{\delta}\\
\frac{a}{\delta} & \mathbf{-}\frac{b}{\delta}%
\end{array}
$} &  & \\
&  & \mathbf{\ddots} & \\
&  &  & \frame{$%
\begin{array}
[c]{rr}%
\mathbf{-}\frac{b}{\delta} & \frac{a}{\delta}\\
\frac{a}{\delta} & \mathbf{-}\frac{b}{\delta}%
\end{array}
$}%
\end{array}
\right]  ;\ \ M_{2}=\left[
\begin{array}
[c]{ccccc}%
0 &  &  &  & \\
& \frame{$%
\begin{array}
[c]{rr}%
-\frac{1}{2d} & \frac{1}{2d}\\
\frac{1}{2d} & -\frac{1}{2d}%
\end{array}
$} &  &  & \\
&  & \ddots &  & \\
&  &  & \frame{$%
\begin{array}
[c]{rr}%
-\frac{1}{2d} & \frac{1}{2d}\\
\frac{1}{2d} & -\frac{1}{2d}%
\end{array}
$} & \\
&  &  &  & 0
\end{array}
\right]  .
\]
We first note that (\ref{lam1eq}) is equivalent to 
\begin{equation}
Mv=-\mu\left(  \frac{\sqrt{2}}{L-l}\right)  v \,, \label{mdecom}
\end{equation}
where $v$ is an eigenvector corresponding to $\sigma$ and $\mu=\lambda
_{1}+\beta_{0}$. Let $v_k$ be the $k^{\mbox{th}}$ component of $v$. Then,
upon using $\delta=b^2-a^2$, we calculate the inner product as
\bsub \label{m1decom}
\begin{align}
 \bar{v}^{t}M_{1}v  &=-\frac{b}{\delta}\left(  \left\vert v_{1}\right\vert
^{2}+\left\vert v_{2}\right\vert ^{2} + \cdots +
\left\vert v_{2K-1}\right\vert^{2}+\left\vert v_{2K}\right\vert ^{2}\right)
  +\frac{a}{\delta}\left( v_{1}\overline{v_{2}}+v_{2}\overline{v_{1}}
 + \cdots + v_{2K-1}\overline{v_{2K}}+v_{2K}\overline{v_{2K-1}}\right)\,, \\
 &=-\frac{1}{2(b-a)} \left( \left\vert v_{1}-v_{2}\right\vert^{2} + \cdots
 + \left\vert v_{2K-1}-v_{2K}\right\vert ^{2} \right) 
-\frac{1}{2(b+a)} \left( \left\vert v_{1}+v_{2}\right\vert^{2} + \cdots
 + \left\vert v_{2K-1}+v_{2K}\right\vert ^{2} \right) \,, \\
 & = -\frac{C_2}{u_{o}\left(l\right)  } - \frac{C_1}{u_{e}\left(l\right)} \,.
\end{align}
\esub
Here and below $C_{i}$ denotes a non-negative constant that may
change from line to line. Similarly, we obtain
\bsub  \label{m2decom}
\begin{align}
\bar{v}^t M_2 v & = -\frac{1}{2d} 
\left( \left\vert v_{2}-v_{3}\right\vert^{2} + \cdots
 + \left\vert v_{2K-2}-v_{2K-1}\right\vert ^{2} \right) \,,\\
    &=-C_{3} \,.
\end{align}
\esub
Upon premultiplying (\ref{mdecom}) by $\bar{v}^{t}$, and using
(\ref{m1decom}) and (\ref{m2decom}), we then divide by $\mu$ to obtain%
\[
 \frac{C_1}{\mu u_{e}\left(  l\right)  }+\frac{C_2}{\mu u_{o}\left(
l\right)  }+\frac{C_{3}}{\mu}=C_{4} \,.
\]
This equation can be rewritten as
\begin{equation}
C_{1}\overline{\mu u_{e}\left(  l\right)  }+C_{2}\overline{\mu u_{o}\left(
l\right)  }+C_{3}\bar{\mu}=C_{4} \,. \label{10apr7:32}%
\end{equation}
From the expressions for $C_i$ in (\ref{m1decom}) and (\ref{m2decom})
it follows that at least one of the $C_1$, $C_2$, or $C_3$ is strictly
positive for any $v\neq 0$.  Next, we return to the equation for
$u_{o},$%
\[
Du_{o}^{\prime\prime}-\mu\frac{u^{\prime}}{w^{\prime}}u_{o}=0 \,; \qquad
u_{o}\left(  0\right)     =0\,, \quad u_{o}^{\prime}\left(  l\right)  =1 \,.
\]
We multiply this equation by $\overline{u_{o}}$ and integrate the
resulting expression by parts to get
\[
\overline{u_{o}}\left(  l\right)  = \int_{0}^{l}
 \left\vert u_{o}^{\prime}\right\vert^{2} \, dx +\frac{\mu}{D}\int_{0}^{l}
 \frac{u^{\prime}}{w^{\prime}}\left\vert u_{o}\right\vert ^{2} \, dx \,.
\]
Multiplying this expression by $\overline{\mu}$ we obtain
\[
\overline{\mu u_{o}\left(  l\right)  }=\bar{\mu}B_{5}+B_{6} \,.
\]
In a similar way, we derive
\[
\overline{\mu u_{e}\left(  l\right)  }=\bar{\mu}B_{7}+B_{8} \,.
\]
Here $B_5$, $B_6$, $B_7$ and $B_8$ are strictly positive constants.
Substituting these expressions into (\ref{10apr7:32}) we conclude that
\[
\bar{\mu} \left(C_1 B_5 + C_2 B_7 + C_3\right)=B \,,
\]
Here $B$ is a real constant, and we note that $C_1B_5+C_2B_7+C_3$ is
strictly positive for any $v\neq 0$.  Finally, by taking the imaginary
part of this expression, we get $\operatorname{Im}\left( \mu\right)
=\operatorname{Im}\lambda_{1}=0.$ This concludes the proof of Theorem
\ref{thm:stability}.

\bigskip

\setcounter{equation}{0}
\setcounter{section}{3}
\section{{Discussion}}{ \label{sec:discuss} }

 In \cite{kew} the one-dimensional Brusselator model was analyzed with the 
following scaling,
\begin{equation*}
\tau_{k}u_{t}    =\eps_{k}D_{k}u_{xx}+\eps_{k}A_{k}
+u^{2}v-\left(  B_{k}+\eps_{k}\right)  u\,, \qquad
v_{t}    =\eps_{k}D_{k}v_{xx}+B_{k}u-u^{2}v \,,
\end{equation*}
on the interval $x\in [0,1]$ . The assumptions on the parameters were
that $\eps_{k}D_{k}\ll 1$, $D_{k}\gg 1$, $A_k=O(1)$,
and $B_k=O(1)$. This model is equivalent to (\ref{bruv}) after the
change of variables $u=a\hat{u}$, $v=a\hat{v}$, $t=
\frac{\tau_{k}}{a^{2}}\hat{t}$, with $a=\sqrt{B_{k} + \eps_{k}}$, and after 
dropping the hat notation. The parameters are mapped to
\[
\eps=\sqrt{\frac{\eps_{k}D_{k}}{B_{k}+\eps_{k}}}\,;
 \qquad \beta_{0}=\frac{\sqrt{B_{k}+\eps_{k}}}{A_{k}}\,; \qquad D=D_{k}%
\beta_{0}\,; \qquad \tau=\frac{1}{\tau_{k}} \,.
\]
One of the main results in \cite{kew} was that a $K$-mesa configuration 
with $K\geq 2$ is unstable only when
\begin{equation}
\label{20jan5:35}D_{k}>\frac{1}{K^{2}}D_{kc}\text{ \ where \ }D_{kc}%
\sim\left\{
\begin{array}
[c]{c}%
\frac{A_{k}^{2}}{2\eps_{k}\ln^{2}\left(  \frac{12\sqrt{2}A_{k}%
B_{k}^{3/2}}{\eps_{k}\left(  \sqrt{2B_{k}}-A_{k}\right)  ^{2}}\right)
},\ \ \ \ 2A_{k}^{2}<B_{k}\,, \ \\
\frac{\left(  \sqrt{2B_{k}}-A_{k}\right)  ^{2}}{2\eps_k\ln^{2}\left(
\frac{12\sqrt{2}}{\eps_{k}A_{k}}B_{k}^{3/2}\right)  },\ \ \ \ 2A_{k}%
^{2}>B_{k} \,.%
\end{array} 
 \right.
\end{equation}
For $D_k$ above this stability threshold a coarsening phenomenon was
observed in \cite{kew}. This process resulted in the annihilation of
some mesas over an exponentially long time scale, until eventually the
number of mesas was decreased sufficiently so that (\ref{20jan5:35})
no longer holds.

The results in this paper together with \cite{kew} provide analytical
bounds on $D$ for the existence and stability of a steady-state $K$-mesa 
pattern. When $\beta_{0}>\sqrt{2},$ (\ref{20jan5:35}) reduces to
\[
D>\frac{\left(  \sqrt{2}\beta_{0}-1\right)  ^{2}}{12\sqrt{2}\beta_{0}%
}\eps^{2}\exp\left(  \frac{1}{\sqrt{2}K\beta_{0}\eps}\right)  .
\]
This provides an exponentially large upper bound on $D$ for the
stability of $K$ mesas.  Roughly speaking, the stability of $K$-mesas when
$\tau$ is sufficiently small is guaranteed on the range
\[
O\left(  \frac{1}{K^{2}}\right)  \ll D \ll O\left(  \eps^{2}
\exp\left(  \frac{1}{\sqrt{2}K\beta_{0}\eps}\right)  \right)  .
\]
If $D$ exceeds an exponentially large upper bound, then the number of
mesas is diminished through a coarsening process. Alternatively, if
$D$ is too small, then self-replication is observed until such time that $D
K^{2}$ is large enough.

{ \begin{figure}[tb]
{ \begin{minipage}[t]{0.48\textwidth}
\begin{center}{
\setlength{\unitlength}{1\textwidth} \begin{picture}(1,1)(0,-0.04)
\put(-0.1,0.03){\includegraphics[width=1.2\textwidth, height=0.9\textwidth]
{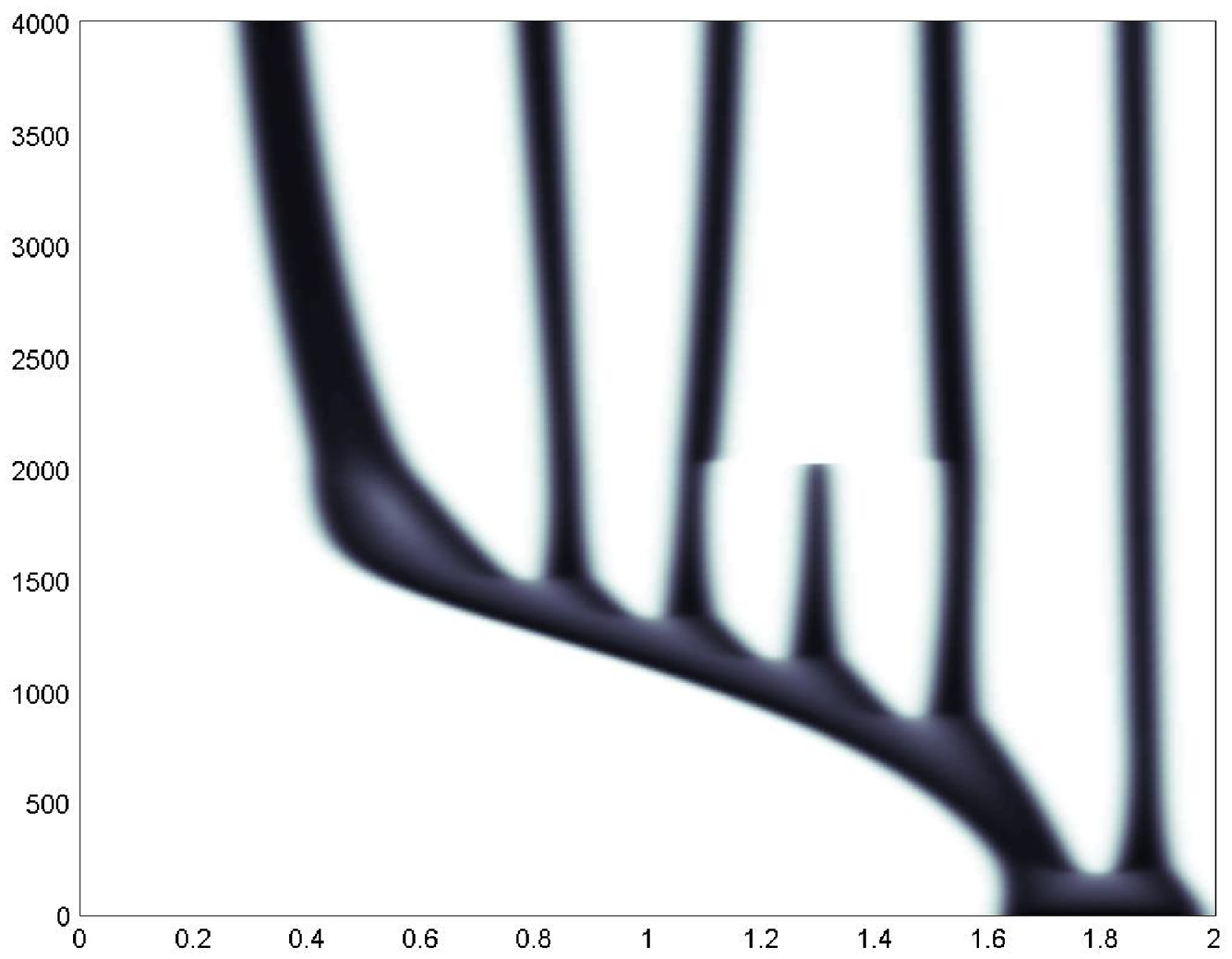}}
\put(0.47,0.00){(a)}
\end{picture}
}\end{center}
\end{minipage}
\begin{minipage}[t]{0.48\textwidth}
\begin{center}{
\setlength{\unitlength}{1\textwidth} \begin{picture}(1,1)(0,-0.04)
\put(-0.1,0.03){\includegraphics[width=1.2\textwidth, height=0.9\textwidth]
{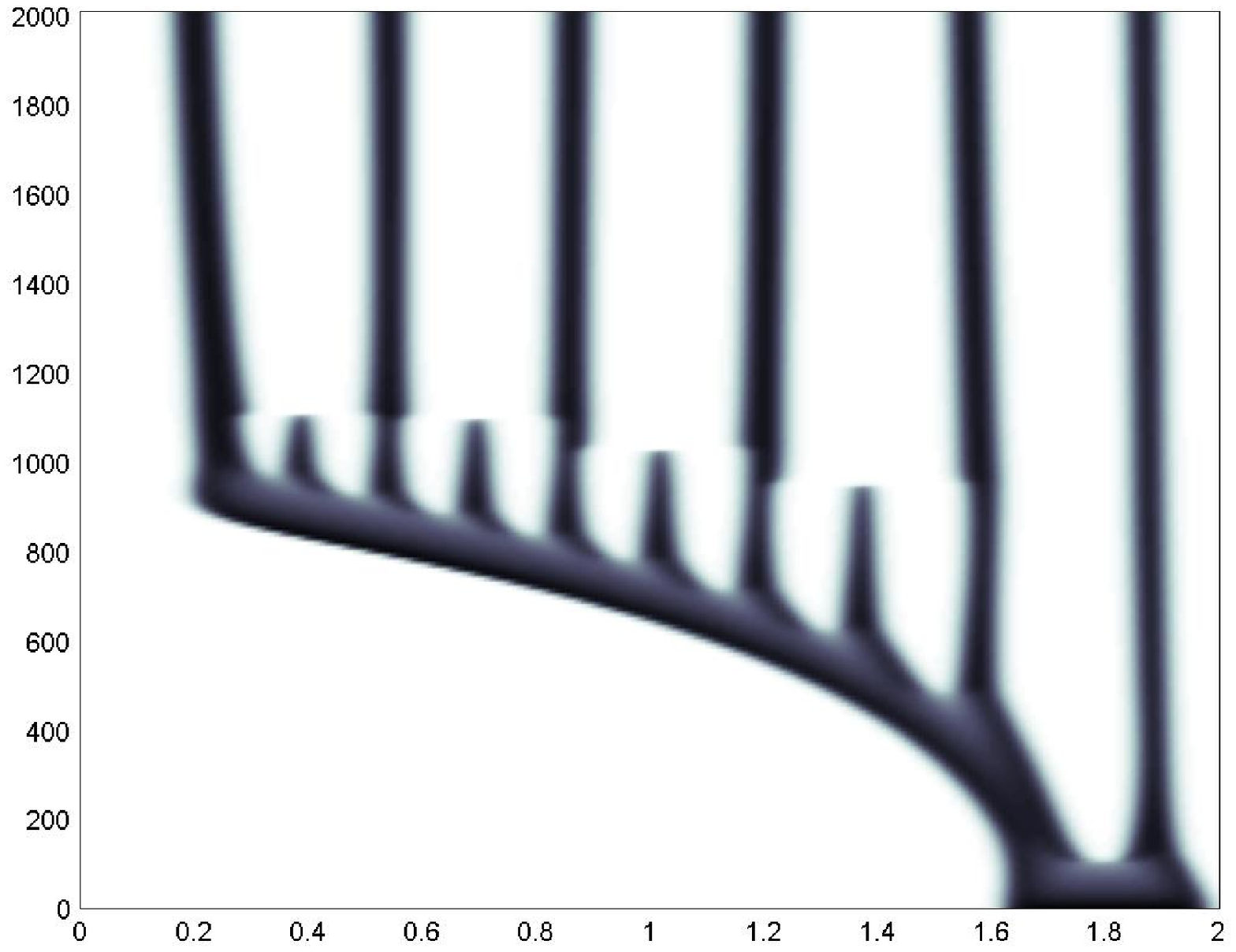}}
\put(0.47,0.00){(b)}
\end{picture}
}
\end{center}
\end{minipage}
}
\caption{
Space-time plots showing
self-replication of a moving mesa. (a) $D=0.1$; (b) $D=0.05$. The other
parameter values are fixed at $\eps = 0.01$, $\beta_0 = 4$, $\tau=0.1$.
Critical threshold values for $K$ mesas are $D_1=0.65$, $D_2=0.16$,
$D_3=0.07$, $D_4=0.04$.
}
\label{fig:move}
\end{figure}}

There are several open problems. 
When the initial data is non-symmetric, it is possible to observe a
sequence of self-replication events without changing parameter values. An
example of this phenomenon is shown in Figure \ref{fig:move}. 
In this case, the
motion of the mesa 
itself causes successive replication events until a steady 
stable state is
reached. 
The minimum number of splitting events is given by
$K=\sqrt{D_c/D}$ where $D_c$ is given by Proposition \ref{prop:Dc}, although
numerical simulations indicate that the eventual number of mesas is
higher than this minimum. Note also that a sequence of
``abortive'' splittings is observed; we speculate that these are
connected to the overcrowding phenomenon as described in \cite{kew},
although the details are unclear. The question of how the motion of the
mesa affects the splitting behaviour is also open.

Another open problem is to study the
stability and dynamics of equilibrium and quasi-equilibrium mesa
patterns when $\tau> 0$. In \cite{kew}, it was shown that when $D
\gg1$, there is a Hopf bifurcation that occurs for $\tau\sim 1.$
However, the analysis there relied on explicit analytical calculations
of the small eigenvalues, which is not possible when
$D=O(1)$. Moreover, it was shown in \cite{kew}, under some additional
conditions, that a Hopf bifurcation can occur leading to a
breather-type instability whereby the center of the mesa remains
stationary, but its width slowly oscillates in time. For $D$ above the
self-replication threshold of a steady-state mesa, we suggest that
such a breather-type instability for $\tau$ sufficiently large can
trigger a dynamic mesa self-replication event if the time-oscillating
mesa plateau width exceeds its maximum steady-state value. Such a
triggering mechanism is explored for a reaction-diffusion system with
piecewise-linear kinetics in \cite{ho3}.  In addition, when $D=O(1)$,
some numerical simulations (not shown) suggest that an oscillatory
traveling-wave instability is also possible, whereby the position of
the center of the mesa oscillates in time, while its width remains
constant.

The second area of open problems is to extend the study of the
existence and stabilty of mesa patterns to two or higher
dimensions. Numerical simulations suggest a slew of possible
patterns. One possibility is to study radially symmetric patterns in a 
ball. One can then obtain a blob-like pattern. Self-replication of
such a pattern can occur as $D$ is decreased sufficiently, and we
expect the core problem in two dimensions to be (\ref{core}) with $y$
replaced by $|y|$. The study of mesa blob-type patterns in an
arbitrary two-dimensional domain is also open. Another possibility is
to trivially extend the one-dimensional mesa pattern into the second
dimension. The resulting mesa-stripe pattern can exhibit transverse
instabilities. This stability problem was examined in
\cite{kww-stripe} for the Gierer-Meinhardt model with saturation 
in the near-shadow limit where mesa self-replication does not
occur. A similar stability analysis in the mesa self-replication regime
is open. The analysis of these and related problems in two dimensions is
the subject of future work.

\section*{Acknowledgements} 
T.~K. was supported by an NSERC Postdoctoral Fellowship, and is grateful
for the hospitality of the Chinese University of Hong Kong where this
work was initiated. M.~J.~W. acknowledges the grant support of NSERC, and
J.~W. thanks the support of RGC of Hong Kong.

\appendix
\newcommand{\newsection}[1]{{\setcounter{equation}{0}}\section{#1}}
\renewcommand{\theequation}{\Alph{section}.\arabic{equation}}
\newsection{The SLEP Reduction}

\textbf{Proof of Lemma \ref{thm:slep-reduction}.} We label the interface
locations by
\[
x_{1-}<x_{1+}<x_{2-}<x_{2+}<\cdots<x_{K-}<x_{K+} \,, %
\]
as illustrated on Fig.~\ref{fig:ss}(b). We then define $l$ and $d$ by
\[
l=(x_{i+}-x_{i-})/2\,, \qquad d=\left(  L-l\right)  =(x_{(i+1)-}-x_{i+})/2 \,.
\]
By symmetry $l$ and $d$ are independent of $i$. 

In the inner region near $x_{i\pm}$, we introduce the inner variables
\begin{equation}
\phi   =\Phi\left(  y\right)  =\Phi_{0}+\eps\Phi_{1}+\cdots\,, \qquad
\psi   =\Psi\left(  y\right)  =\Psi_{0}+\eps\Psi_{1}+\cdots\,, \qquad
\lambda=\alpha\lambda_{1}+\cdots \,, \qquad
 y=\eps^{-1}(x-x_{i\pm})\,, \label{app:inn}
\end{equation}%
where $\alpha=\eps^2\alpha_0$.  Upon substituting (\ref{app:inn}) into
(\ref{lin}), we obtain the leading-order system
\begin{equation*}
 \Phi_{0}^{\prime\prime}-f_{u}\left(  U_{0},W_{0}\right)  \Phi_{0}%
-f_{w}\left(  U_{0},W_{0}\right)  \Psi_{0} =0 \,, \qquad
 \Psi_{0}^{\prime\prime}=0 \,. %
\end{equation*}
Here $U_0$ and $W_0$, satisfying (\ref{U0}), are given in
(\ref{U0W0}). We take the $+$ sign for $U_{0}$ in (\ref{U0W0}) for the
inner region near $x=x_{i-}$, and the $-$ sign in (\ref{U0W0}) for the
region near $x=x_{i+}$. By differentiating (\ref{U0}) with respect to $y$, 
we get
\begin{equation}
\Phi_{0}=c_{i\pm}U_{0}^{\prime}\left(  y\right)\,, \qquad \Psi_{0}=0 \,,
 \label{app:lead}
\end{equation}
where $c_{i\pm}$, for $i=1,\ldots,K$, are constants to be determined.

Since $\Psi_0\equiv 0$, we obtain the following problem for $\Phi_1$ and 
$\Psi_1$ at next order:
\begin{equation*}
 \Phi_{1}^{\prime\prime}-f_{u}\left(  U_{0},W_{0}\right)  \Phi_{1}-
 f_{w}\left(  U_{0},W_{0}\right)  \Psi_{1} =
  \Phi_{0}\left[  f_{uu}\left(  U_{0},W_{0}\right)  U_{1}+f_{uw}\left(
U_{0},W_{0}\right)  W_{1}\right]  \,, \qquad \Psi_{1}^{\prime\prime}=0 \,.
\end{equation*}
In order to match to the outer solution constructed below we require that
$\Psi_1$ is a constant. We then multiply the equation for $\Phi_1$ by
$U_{0}^{\prime}$, and integrate the resulting expression by parts, to get
\begin{equation}
 \int_{-\infty}^{\infty} U_{0}^{\prime}f_{w}\left(
U_{0},W_{0}\right)  \Psi_{1} \, dy =-\int_{-\infty}^{\infty}
 \Phi_{0}U_{0}^{\prime}\left[ 
  f_{uu}\left(U_{0},W_{0}\right)  U_{1}+f_{uw}\left(  U_{0},W_{0}\right)
  W_{1}\right]  \, dy \,. \label{app:temp1}
\end{equation}
To simplify the right-hand side of (\ref{app:temp1}), we differentiate the
equation for $U_1$ in (\ref{U1}) to get
\begin{equation*}
U_{1}^{\prime\prime\prime}-f_{u}\left(  U_{0},W_{0}\right)  U_{1}^{\prime
} = f_{w}\left(  U_{0},W_{0}\right)  W_{1}^{\prime} + U_{0}^{\prime}\left[
f_{uu}\left(  U_{0},W_{0}\right)  U_{1}+f_{uw}\left(  U_{0},W_{0}\right)
W_{1}\right]  =0 \,.
\]
Upon multiplying this equation by $\Phi_0$, and integrating the resulting
expression by parts, we obtain the identity
\begin{equation}
\int_{-\infty}^{\infty} \Phi_0 U_{0}^{\prime}\left[ 
 f_{uu}\left(  U_{0},W_{0}\right) U_{1}+f_{uw}\left(  U_{0},W_{0}\right) 
 W_{1}\right]\, dy  = -\int_{-\infty}^{\infty} f_{w}(U_0, W_0)
 W_{1}^{\prime}\Phi_{0} \,dy \,. \label{app:temp12}
\end{equation}
In (\ref{app:temp1}), we use (\ref{app:temp12}), 
$\Phi_{0}=c_{i\pm} U_{0}^{\prime}$, and the facts that $\Psi_1$ and 
$W_1^{\prime}$ are constants (see (\ref{w1prime}), to get
\begin{equation}
  \Psi_1 \int_{-\infty}^{\infty} U_{0}^{\prime}f_{w}\left(
U_{0},W_{0}\right)\, dy  =c_{i\pm}W_{1}^{\prime}\int_{-\infty}^{\infty}
 f_{w}(U_0,W_0) U_{0}^{\prime} \, dy \,. \label{10apr7:37}
\end{equation}
Since $f_{w}=-U_{0}^2$ and $\int_{-\infty}^{\infty} U_{0}^2 U_{0}^{\prime}\,
dy\neq 0$, (\ref{10apr7:37}) yields $\Psi_1\equiv c_{i\pm} W_{1}^{\prime}$.
However, $W_{1}^{\prime}=w^{\prime}(x_{i\pm})$ from (\ref{w1prime}), and 
$\Psi_1=\eps \psi(x_{i\pm})$, where $\psi(x)$ is the outer solution for
(\ref{lin}). Therefore, we have the following key relationship:
\begin{equation}
   \psi(x_{i\pm}) =  \eps c_{i\pm} w^{\prime}(x_{i\pm}) \,. \label{app:key}
\end{equation}

Next, we derive an outer equation for $\psi$. In the outer region, defined
on the union of the subintervals $-KL<x<x_{1-}$, $x_{i-}<x<x_{i+}$ for
$i=1,\ldots,K$, and $x_{k+}<x<KL$, we obtain from (\ref{lin}) the 
leading-order system
\begin{equation*}
 \phi   = -\frac{f_{w}}{f_{u}}\psi=\frac{u^{\prime}}{w^{\prime}}\psi 
\,, \qquad \lambda_{1}\phi   = D\psi_{xx}-\beta_{0}\phi \,.
\end{equation*}
These equations can be combined to give
\[
D\psi_{xx}-\left(  \lambda_{1}+\beta_{0}\right)  \frac{u^{\prime}}{w^{\prime}%
}\psi=0 \,.
\]
To determine the jump condition for $\psi$ across $x=x_{i\pm}$, we use
use the inner result $\phi\sim c_{i\pm} U_{0}^{\prime}$ to derive
\begin{equation}
 \left[D\psi^{\prime}\right]|_{i\pm}\equiv D\psi^{\prime}(x_{i\pm}^{+})-
D\psi^{\prime}(x_{i\pm}^{-})= c_{i\pm} (\lambda_1+\beta_0)
 \int_{-\infty}^{\infty} 
 U_{0}^{\prime}\, dy = \mp \sqrt{2} \eps c_{i\pm} 
 \left(\lambda_1 + \beta_0\right) \,. \label{app:jump}
\end{equation}
Therefore, the outer problem for $\psi$ on $-KL<x<KL$ is
\begin{equation}
D\psi^{\prime\prime}-\left(  \beta_{0}+\lambda_{1}\right) \frac{u^{\prime}}
 {w^{\prime}} \psi=-\sqrt{2}\eps\left(  \beta_{0}+\lambda_{1}\right) 
 \left( \sum_{i=1}^{K} \left[c_{i+}\delta\left( x- x_{i+}\right)  -c_{i-}\delta
 \left( x- x_{i-}\right) \right]\right)  \,, \label{10apr7:43}%
\end{equation}
with $\psi^{\prime}(\pm KL)=0$. Note that in those outer regions where
$u=0$ to leading order, we have $\frac{u^{\prime}}{w^{\prime}}=0$. 

\begin{figure}[tb]
\begin{center}
\setlength{\unitlength}{1\textwidth} \begin{picture}(1,0.5)(0,0)
\put(0.03,0.02){\includegraphics[width=0.9\textwidth]{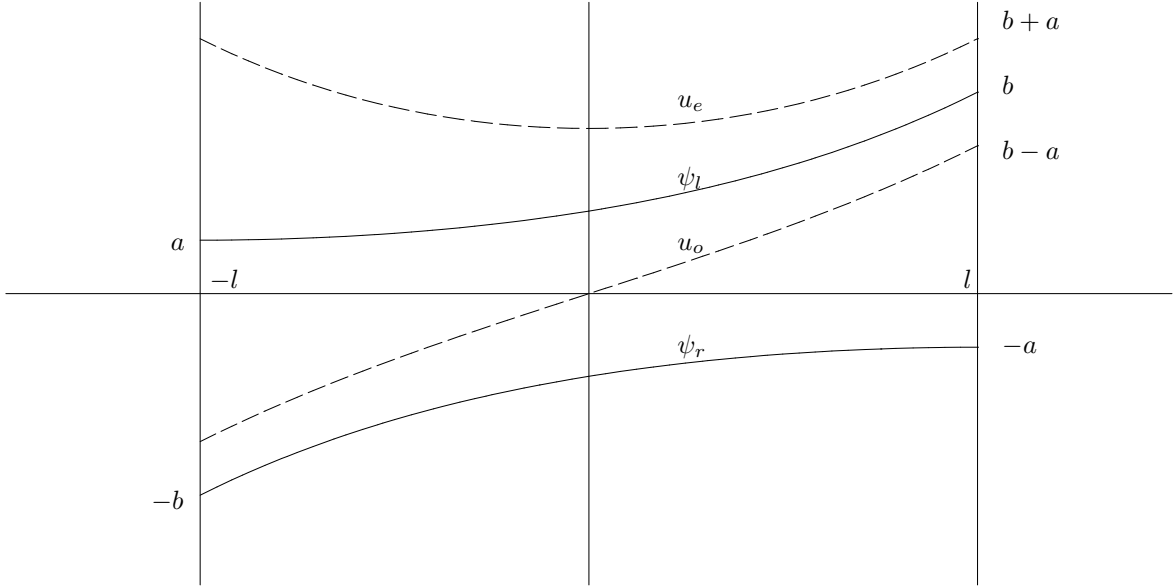}}
\put(0.55,0.20){$\psi_r$}
\put(0.55,0.28){$u_o$}
\put(0.55,0.33){$\psi_l$}
\put(0.55,0.39){$u_e$}
\put(0.77,0.25){$l$}
\put(0.19,0.25){$-l$}
\put(0.8, 0.2){$-a$}
\put(0.8, 0.35){$b-a$}
\put(0.8, 0.4){$b$}
\put(0.8, 0.45){$b+a$}
\put(0.16, 0.28){$a$}
\put(0.145, 0.08){$-b$}
\end{picture}
\end{center}
\caption{Symmetry of $\psi_r$, $\psi_l$, $u_o$ and $u_e$.}%
\label{fig:philr}%
\end{figure}

{\bf Single Mesa:} We first analyze (\ref{10apr7:43}), together with
(\ref{app:key}), for the special case of a single mesa where $K=1$.
For this case $x_{1-}=-l,\ x_{1+}=+l$, and and
$x\in\lbrack-L,L]$. Then (\ref{10apr7:43})\ is equivalent to%
\begin{gather*}
D\psi^{\prime\prime}-\left(  \beta_{0}+\lambda_{1}\right)  \frac{u^{\prime}%
}{w^{\prime}}\psi   =0\,, \quad  x\in(-l,l)\,; \qquad
\psi^{\prime\prime}    =0\,, \quad x\in\left(  l,L\right)  \cup\left(
-L,-l\right)\,;\\
D\psi^{\prime}\left(  l^{+}\right)  -D\psi^{\prime}\left(  l^{-}\right)   
=-\sqrt{2}\eps\left(  \beta_{0}+\lambda_{1}\right)  c_{1+}\,, \qquad
D\psi^{\prime}\left(  -l^{+}\right)  -D\psi^{\prime}\left(  -l^{-}\right)   
=\sqrt{2}\eps\left(  \beta_{0}+\lambda_{1}\right)  c_{1-} \,, 
\end{gather*}
with $\psi^{\prime}\left(  \pm L\right)=0$. By solving for
$\psi$ on $x\in\left(  l,L\right)  \cup\left(-L,-l\right)$, this
system reduces to
\bsub \label{app:reduce}
\begin{gather}
D\psi^{\prime\prime}-\left(  \beta_{0}+\lambda_{1}\right)  \frac{u^{\prime}%
}{w^{\prime}}\psi   =0\,, \quad x\in(-l,l)\,, \\
D\psi^{\prime}\left(  l\right)     =\sqrt{2}\eps\left(  \beta
_{0}+\lambda_{1}\right)  c_{1+} \,; \qquad
D\psi^{\prime}\left(  -l\right)     =\sqrt{2}\eps\left(  \beta
_{0}+\lambda_{1}\right)  c_{1-} \,. %
\end{gather}
\esub
We represent $\psi$ in terms of the solutions $\psi_{l}$ and $\psi_{r}$ to
\bsub \label{app:landr}
\begin{equation}
D\psi^{\prime\prime}-\left(  \beta_{0}+\lambda_{1}\right)  \frac{u^{\prime}
}{w^{\prime}}\psi = 0 \,,  \quad x\in(-l,l) \,, \label{17nov12:52}%
\end{equation}
with either
\begin{equation}
\psi_{l}^{\prime}\left(  -l\right)  =0 \,,\ \ \ \psi_{l}^{\prime}\left(
l\right)  =1 \, \qquad \mbox{or} \qquad
\psi_{r}^{\prime}\left(  -l\right)  =1 \,, \ \ \ \psi_{r}^{\prime}\left(
l\right)  =0 \,. \label{app:landr_bc}
\end{equation}
\esub

We then define $a$ and $b$ by
\bsub \label{app:andb}
\begin{equation}
a\equiv\psi_{l}\left(  -l\right) \,, \qquad b\equiv\psi_{l}\left(  l\right)\,.
\end{equation}
Since $\frac{u^{\prime}}{w^{\prime}}$ is an even function, we also have
\begin{equation}
\psi_{r}\left(  -l\right)  =-b\,,\qquad \psi_{r}\left(  l\right)  =-a \,.
\end{equation}
\esub
In terms of $\psi_l$ and $\psi_r$, the solution for $\psi$ is
$\psi=A_{l}\psi_{l}+A_{r}\psi_{r}$, where
$\psi^{\prime}\left(  -l\right)  =A_{r}$ and $\psi^{\prime}\left(  +l\right)
=A_{l}$. By satisfying the boundary conditions for $\psi$ in
(\ref{app:reduce}), we get
\[
A_{l}=\frac{\sqrt{2}\eps}{D} \left(  \beta_{0}+\lambda_{1}\right)
c_{1+}\,, \qquad A_{r}=\frac{\sqrt{2}\eps}{D} \left(  \beta_{0}+\lambda
_{1}\right)  c_{1-} \,.
\]
In terms of this solution, we write the matrix system
\begin{equation}
\left[
\begin{array}
[c]{c}%
\psi\left(  l\right) \\
\psi\left( -l\right)
\end{array}
\right]     =\left[
\begin{array}
[c]{cc}%
\psi_{l}\left(  l\right)  & \psi_{r}\left(  l\right) \\
\psi_{l}\left(  -l\right)  & \psi_{r}\left(  -l\right)
\end{array}
\right]  \left[
\begin{array}
[c]{c}%
A_{l}\\
A_{r}%
\end{array}
\right] =\frac{\sqrt{2}\eps}{D}\left(  \beta_{0}+\lambda_{1}\right)
\left[
\begin{array}
[c]{cc}%
b & a\\
a & b
\end{array}
\right]  \left[
\begin{array}
[c]{c}%
c_{1+}\\
-c_{1-}%
\end{array}
\right] \,. \label{app:mat1}
\end{equation}

To calculate an independent expression for $\psi(\pm l)$ we use the
identity (\ref{app:key}), which states $\psi(\pm l)=c_{1\pm}\eps
w^{\prime}(\pm l)$. To calculate $w^{\prime}(\pm l)$, we recall that
$Dw^{\prime\prime}=-1$ for $x\in (-L,-l)$ and for $x\in (l,L)$. With
$w^{\prime}(\pm L)=0$, this gives  $w^{\prime}(\pm l)=\pm(L-l)/D$. Therefore,
\begin{equation}
\left[
\begin{array}
[c]{c}%
\psi\left(  l\right) \\
\psi\left(  -l\right)
\end{array}
\right]  =\frac{(L-l)}{D}\eps\left[
\begin{array}
[c]{c}%
c_{1+}\\
-c_{1-}%
\end{array}
\right]  \,. \label{app:mat2}
\end{equation}
Combining (\ref{app:mat1}) and (\ref{app:mat2}), we get
\[
\left[
\begin{array}
[c]{cc}%
b & a\\
a & b
\end{array}
\right]  \left[
\begin{array}
[c]{c}%
c_{1+}\\
-c_{1-}%
\end{array}
\right]  =\frac{(L-l)}{\sqrt{2}}\frac{1}{\left(  \beta_{0}+\lambda_{1}\right)
}\left[
\begin{array}
[c]{c}%
c_{1+}\\
-c_{1-}%
\end{array}
\right]  .
\]
The eigenvalues of the matrix on the left-hand side of this expression
are $b\pm a$. Therefore, $\lambda_{1}$ must satisfy
\begin{equation}
\frac{L-l}{\sqrt{2}}\frac{1}{\left(  \beta_{0}+\lambda_{1}\right)  }=b\pm a.
\label{5dec9:40}
\end{equation}
Finally, we rewrite (\ref{5dec9:40}) in terms of new functions
$u_o(x)$ and $u_e(x)$ defined by
\begin{equation}
u_{o}\equiv \psi_{l}+\psi_{r}\,, \qquad u_{e}=\psi_{l}-\psi_{r} \,. 
 \label{uoue}
\end{equation}
Then, $u_{o}$ is odd and $u_{e}$ is even, and both satisfy
(\ref{17nov12:52}) with the side conditions%
\begin{equation*}
u_{o}\left(  0\right) =0,\ \ \ u_{o}^{\prime}\left(  l\right)  =1\,;
 \qquad u_{e}^{\prime}\left(  0\right)  =0,\ \ \ u_{e}^{\prime}\left(l\right)
=1 \,.
\end{equation*}
Moreover, by using (\ref{app:andb}) and (\ref{uoue}), we calculate
\[
u_{o}\left(l\right)  =b-a\,,\qquad u_{e}\left(l\right)  =b+a \,.
\]
Therefore, from (\ref{5dec9:40}), the eigenvalues must satisfy
\begin{equation*}
\frac{(L-l)}{\sqrt{2}}\frac{1}{\left(  \beta_{0}+\lambda_{1}\right)  }%
=u_{e}\left(  l\right) \,, \qquad \mbox{or} \qquad
\frac{(L-l)}{\sqrt{2}}\frac{1}{\left(  \beta_{0}+\lambda_{1}\right)  }%
=u_{o}\left(  l\right)  \,. 
\end{equation*}
The corresponding eigenfunctions are either even or odd. This proves
(\ref{eig:onemesa}) of Lemma \ref{thm:slep-reduction}.

{\bf General case:} We now consider the case of $K$ mesas with
$K>1$. On each subinterval we solve for $\psi$ to obtain
\begin{gather*}
\psi   =A_{il}\psi_{li}+A_{ir}\psi_{ri} \,, \qquad x\in\lbrack x_{i-}%
,x_{i+}] \,, \\
\psi   =C_{i}+D_{i}\left(  x-x_{i+}\right)\,, \qquad x\in\lbrack x_{i+},
 x_{(i+1)-}] \cup \lbrack -KL,x_{1-}] \cup \lbrack x_{K+}, KL] \,,
\end{gather*}
where the coefficients $A_{il}$, $A_{ir}$, $C_i$, and $D_i$ are to be
found.  The functions $\psi_{l},$ $\psi_{r}$ solve
$D\psi^{\prime\prime}-\left( \beta _{0}+\lambda_{1}\right)
\frac{u^{\prime}}{w^{\prime}}\psi=0$ with
\begin{equation*}
\psi_{li}^{\prime}\left(  x_{i-}\right)     =0,\ \ \ \psi_{li}^{\prime
}\left(  x_{i+}\right)  =1 \,; \qquad
\psi_{ri}^{\prime}\left(  x_{i-}\right)     =1,\ \ \ \psi_{ri}^{\prime
}\left(  x_{i+}\right)  =0.
\end{equation*}
Similar to the case of a single mesa, we have
\begin{equation*}
\psi_{li}\left(  x_{i-}\right)   =a \,, \qquad \psi_{li}\left(  x_{i+}\right)
=b \,, \qquad \psi_{ri}\left(  x_{i-}\right)   =-b \,, \qquad
  \psi_{ri}\left(  x_{i+}\right) =-a \,.
\end{equation*}

By satisfying the jump condition for $D\psi^{\prime}$ across $x=x_{i\pm}$ from
(\ref{10apr7:43}) we obtain
\bsub \label{app:kjump}
\begin{align}
D\left(\psi^{\prime}\left(  x_{i+}^{+}\right)  -\psi^{\prime}\left(  x_{i+}%
^{-}\right) \right)  &  =D_{i}-A_{il}=-\sqrt{2}\eps\left(  \beta_{0}%
+\lambda_{1}\right)  c_{i+}\,, \\
D\left( \psi^{\prime}\left(  x_{i-}^{+}\right)  -\psi^{\prime}\left(  x_{i-}%
^{-}\right)\right) &  =A_{ir}-D_{i-1}=\sqrt{2}\eps\left(  \beta_{0}%
+\lambda_{1}\right)  c_{i-} \,. 
\end{align}
\esub
Then, by the continuity of $\psi$ across $x_{i\pm}$ we get
\begin{equation*}
aA_{il}-bA_{ir}    =C_{i-1}+ 2 D_{i-1}d \,, \qquad bA_{il}-aA_{ir}  =C_{i} \,,
\end{equation*}
where $2d=x_{i-}-x_{\left(  i-1\right)+}$. Then, by using 
$\psi^{\prime}\left(\pm KL\right)=0$, we solve for $C_{i}$ and $D_{i}$ to 
obtain
\begin{equation}
D_{0}  = D_{K}=0 \,, \qquad C_{i}  =bA_{il}-aA_{ir} \,, \qquad
D_{i}    =\frac{1}{2d}\left( aA_{\left(  i+1\right)  l}-
 bA_{\left(  i+1\right)r} -bA_{il}+aA_{ir} \right) \,. \label{app:coeff}
\end{equation}
Moreover, we calculate
\[
\psi\left(  x_{i-}\right)  =A_{il}a-A_{ir}b \,, \qquad
  \psi\left(  x_{i+}\right) =A_{il}b-A_{ir}a \,,
\]
so that
\begin{equation}
A_{il}=\frac{1}{\delta} \left(b\psi\left(  x_{i+}\right)  -a\psi\left(
  x_{i-}\right)\right)\,, \qquad A_{ir}=\frac{1}{\delta} \left(
 a\psi\left(  x_{i+}\right)  -b\psi\left(x_{i-}\right)  \right) \,,
\label{app:coeff1}
\end{equation}
where $\delta \equiv b^{2}-a^{2}$. Therefore, substituting
(\ref{app:coeff1}) and (\ref{app:coeff}) into (\ref{app:kjump}), we
obtain
\begin{gather*}
\frac{\sqrt{2}\eps}{D}\left(  \beta_{0}+\lambda_{1}\right)  c_{i-}    =\frac
{a}{\delta}\psi\left(  x_{i+}\right)  -\frac{b}{\delta}\psi\left(
x_{i-}\right)  -\frac{1}{2d}\psi\left(  x_{i-}\right)  +\frac{1}{2d}%
\psi\left(  x_{(i-1)+}\right)  \,, \quad 1<i\leq K \,, \\
-\frac{\sqrt{2}\eps}{D} \left(  \beta_{0}+\lambda_{1}\right)  c_{i+}  
=\ \frac{a}{\delta}\psi\left(  x_{i-}\right)  -\frac{b}{\delta}\psi\left(
x_{i+}\right)  -\frac{1}{2d}\psi\left(  x_{i+}\right)  +\frac{1}{2d}%
\psi\left(  x_{(i+1)-}\right) \,, \quad 1\leq i<K \,, \\
\frac{\sqrt{2}\eps}{D}\left(  \beta_{0}+\lambda_{1}\right)  c_{1-}   =\frac
{a}{\delta}\psi\left(  x_{1+}\right)  -\frac{b}{\delta}\psi\left(
x_{1-}\right)\,, \qquad
-\frac{\sqrt{2}\eps}{D} \left(  \beta_{0}+\lambda_{1}\right)  c_{K+}    =\frac
{a}{\delta}\psi\left(  x_{K-}\right)  -\frac{b}{\delta}\psi\left(
x_{K+}\right)  \,. 
\end{gather*}
This system can be written in matrix form as
\begin{equation}
\sqrt{2}\eps\left(  \beta_{0}+\lambda_{1}\right)  v=M z \,, \qquad
v \equiv \left[
\begin{array}
[c]{c}%
c_{1-}\\
-c_{1+}\\
\vdots\\
c_{K-}\\
-c_{K+}%
\end{array}
\right]  \qquad z \equiv \left[
\begin{array}
[c]{c}%
\psi\left(  x_{1-}\right) \\
\psi\left(  x_{1+}\right) \\
\vdots\\
\psi\left(  x_{K-}\right) \\
\psi\left(  x_{K+}\right)
\end{array}
\right]  .
\label{11apr11:32}%
\end{equation}
Here the matrix $M$ is given in (\ref{M}). Finally, we use (\ref{app:key})
to calculate $\psi(x_{i\pm})=\pm c_{i\pm}{(L-l)/D}$. This yields that
$z=\eps(l-L){v/D}$. Therefore, (\ref{11apr11:32}) becomes
\begin{equation}
   M v = \frac{\sqrt{2}}{l-L} (\beta_0+\lambda_1) v \,. \label{app:final}
\end{equation}
The eigenvalue problem (\ref{app:final}) is equivalent to that in
Lemma \ref{thm:slep-reduction}. The eigenvalues of $M$ were calculated
explicitly in \cite{rw}, and are given by (\ref{sigma}). This completes the
proof of Lemma \ref{thm:slep-reduction}.

\end{document}